\documentclass[12pt]{article}
\usepackage{latexsym,amsmath,amsthm,amssymb,amsfonts,amsbsy,wasysym,stmaryrd,hyperref}
\usepackage[dvips]{graphicx,epsfig}
\usepackage{color}
\usepackage{url}

\usepackage{verbatim}

\usepackage{bbold}

\usepackage{setspace}

\parskip=\medskipamount

\csname @addtoreset\endcsname{equation}{section}

\usepackage{empheq}

\def\Real{{\mathbb R}}
\def\Comp{{\mathbb C}}

\def\1{1\hspace{-4pt}1}
\def\j1{\widetilde{1\hspace{-4pt}1}}

\def\nequi1{\equiv{\hspace{-4pt}1}}

\def\bec{\begin{center}}
\def\ec{\end{center}}
\def\a{\alpha}
\def\ad{\dot{\a}}

\def\e{\epsilon}

\def\s{\sigma}
\def\S{\Sigma}

\def\nn{\nonumber}
\newcommand{\eq}[1]{(\ref{#1})}

\def\ed{\end{document}}
\newcommand{\w}[1]{\\[0.#1cm]}
\def\be{\begin{equation}}
\def\ee{\end{equation}}
\def\bea{\begin{eqnarray}}
\def\eea{\end{eqnarray}}
\def\ba{\begin{array}}
\def\ea{\end{array}}

\def\ad{\dot\alpha}

\def\wt{\widetilde}

\definecolor{rougef}{rgb}{0.7,0,0}
\definecolor{vertf}{rgb}{0,0.6,0}
\definecolor{bleuf}{rgb}{0,0,0.9}

\begin{document}

\begin{center}


{\Large\bf Frobenius-Chern-Simons gauge theory}


\vskip .5cm

{\bf Roberto Bonezzi\,$^1$, Nicolas Boulanger\,$^1$, 
Ergin Sezgin\,$^2$ and Per Sundell\,$^3$ }\\

\vskip .5cm

{\small{
{\em $^1$ \hskip -.1truecm
\textit{Groupe de M\'ecanique et Gravitation, Physique Th\'eorique et Math\'ematique,\\ 
Universit\'e de Mons -- UMONS, 20 Place du Parc, B-7000 Mons, Belgium}
\vskip 1pt }

{\small{\tt nicolas.boulanger@umons.ac.be, roberto.bonezzi@umons.ac.be}}
}}

\vskip 10pt

{\small{
{\em $^2$ \hskip -.1truecm George and Cynthia Woods Mitchell Institute for Fundamental
Physics and Astronomy \\ Texas A\& M University, College Station,
TX 77843, USA
\vskip 1pt }

{\small{\tt sezgin@tamu.edu}}
}}

\vskip 10pt

{\small{
{\em $^3$ \hskip -.1truecm
\textit{Departamento de Ciencias F\'isicas, Universidad Andres Bello, Republica 220, Santiago de Chile}
\vskip 1pt }

{\small{\tt per.anders.sundell@gmail.com}}
}}

\vskip 0.3 cm

{{ABSTRACT}}\\[3ex]
\end{center}

\noindent Given a set of differential forms on an odd-dimensional 
noncommutative manifold valued in an internal associative algebra 
${\cal H}$, we show that the most general cubic covariant Hamiltonian 
action, without mass terms, is controlled by an $\mathbb Z_2$-graded 
associative algebra ${\cal F}$ with a graded symmetric nondegenerate 
bilinear form. 
The resulting class of models provide a natural generalization of the 
Frobenius-Chern-Simons model (FCS) that was proposed in
\href{http://arxiv.org/abs/arXiv:1505.04957}{\texttt{arXiv:1505.04957}}
as an off-shell formulation of the minimal bosonic 
four-dimensional higher spin gravity theory. 
If ${\cal F}$ is unital and the $\mathbb Z_2$-grading is induced from 
a Klein operator that is outer to a proper Frobenius subalgebra,
then the action can be written on a form akin to topological open
string field theory in terms of a superconnection  
valued in ${\cal H}\otimes {\cal F}$. 
We give a new model of this type based on a twisting of
$\Comp[\mathbb Z_2\times \mathbb Z_4]$, which leads to
self-dual complexified gauge fields on $AdS_4$.
If ${\cal F}$ is 3-graded, the FCS model can be 
truncated consistently as to zero-form constraints 
on-shell.
Two examples thereof are a twisting of $\Comp[(\mathbb Z_2)^3]$ 
that yields the original model, and the Clifford algebra ${\cal C}\ell_{2n}$ which 
provides an FCS formulation of the bosonic Konstein--Vasiliev 
model with gauge algebra $hu(4^{n-1},0)\,$.

\pagebreak

\setcounter{page}{1}


\tableofcontents


\section{Introduction}
\label{sec:Section1}

In \cite{Boulanger:2015kfa}, a modified version of Vasiliev's four-dimensional higher 
spin gravity \cite{Vasiliev:1990en} has been introduced, with enlarged gauge symmetry
and a dynamical two-form master field but with the same master zero-form as the original
model.
As a result, the two models propagate the same local degrees of freedom
but the new one has (much) fewer higher spin gauge invariant
observables, which could be an advantage in finding an effective action
along the topological field theory inspired approach proposed and studied in \cite{Colombo:2010fu,Boulanger:2011dd,Sezgin:2011hq,Boulanger:2012bj,Colombo:2012jx}.

More specifically, the enlarged master field content consists of the original Weyl zero-form; 
two one-forms gauging one-sided actions of a complexified higher spin algebra; 
a two-form that contains topological degrees of freedom (including moduli for the
star product algebra on the internal twistor space and fluxes in spacetime); 
corresponding  bulk Lagrange multipliers; and, finally, the master fields 
required for a gapless duality extension of the model \cite{Boulanger:2011dd}. 
The key to the extension is the fact that the Cartan integrability of 
the resulting enlarged system of unfolded equations of motion is
controlled by an an internal eight-dimensional $\mathbb Z_2$-graded Frobenius algebra, 
such that the full field content can be assembled into 
a single flat superconnection valued in a direct product of 
this algebra and the original associative higher spin star 
product algebra.

An action principle can then be constructed following the 
Alexandrov-Kontsevitch-Schwarz-Zaboronsky (AKSZ) procedure
\cite{Alexandrov:1995kv}, by introducing an auxiliary fifth 
commuting dimension and writing a covariant Hamiltonian 
action for the superconnection on the resulting
nine-dimensional base manifold; further references and
a review can be found in \cite{Arias:2016ajh}. 
Formally, the superconnection is an odd element of
an underlying associative superbundle, whose 
superdifferential can be used to write down a 
Chern--Simons-like cubic action\footnote{The complete 
specification of the model requires careful 
choices of classes of symbols on the 
noncommutative twistor spaces on the base 
manifold and the fiber, for which we refer 
to~\cite{Boulanger:2015kfa} and \cite{Iazeolla:2012nf}.},
leading to what shall henceforth refer to as the 
Frobenius-Chern-Simons (FCS) formulation of four-dimensional 
higher spin gravity, or the minimal FCS gauge theory.

The mathematical structure of the FCS model suggests a number 
of generalizations worthy of investigation, in particular in 
relation to the proposed relationship
between (massless) higher spin gravity and topological open 
strings~\cite{Engquist:2005yt}, 
later verified directly at the level of amplitudes 
in~\cite{Colombo:2012jx,Didenko:2012tv}.

In this paper we shall examine the structure of the most general FCS 
model consisting of a set of even and odd forms on an 
odd-dimensional noncommutative base manifold valued in an internal 
associative algebra ${\cal H}$ and with canonical kinetic terms 
and general cubic Hamiltonian without mass terms. 
As we shall see, the gauge symmetry of the action, or equivalently, 
the Cartan integrability of the equations of motion, results in an 
action for a superconnection 
valued in ${\cal H}\otimes {\cal F}$ where ${\cal F}$ is a 
$\mathbb Z_2$-graded
associative algebra with a graded symmetric nondegenerate bilinear 
form, that we refer to as a $\mathbb Z_2$-graded quasi-Frobenius 
algebra, as it does not have to contain a unity.
In general, the resulting field equations may contain  
integrable zero-form constraints, which can be treated
within the AKSZ scheme.

We then focus on unital algebras in which the $\mathbb Z_2$-grading 
is generated by a  Klein operator that is outer to a proper 
Frobenius subalgebra ${\cal F}_0\subset {\cal F}$. 
Simple examples of these generalized FCS models, that we shall 
present below, are based on matrix algebras and 
twisted group algebras \cite{conlon1964twisted,serre2005groupes}. 
In this category, we shall present a simple model based on 
an eight-dimensional Frobenius algebra that leads to
a variant of the original FCS model with a zero-form
constraint, containing a branch consisting of self-dual 
complexified gauge fields along the lines of \cite{Iazeolla:2007wt}.

A subset of the $\mathbb Z_2$-graded models exhibit a 
refined 3-grading that can used to truncate the top-forms 
consistently together with some of the zero-forms and 
next-to-top forms, as to obtain a subclass of FCS models 
without zero-form constraints.
We shall describe a specific truncation scheme,
that employs an inner Klein operator in defining the 
3-grading, and provide examples thereof based on 
matrix algebras, Clifford algebras and twisted group 
algebras.
In particular, the original FCS model arises within
this subclass from a twisting the $(\mathbb Z_2)^3$ algebra.
Another set of examples based on matrix and Clifford algebras
furnish novel off-shell formulations of a class of bosonic 
Konstein--Vasiliev \cite{Konshtein:1988yg,Konstein:1989ij} 
models, that differ from their direct FCS extensions (based
on the direct product of the Frobenius algebra of 
minimal FCS model and the Konstein--Vasiliev matrix algebra).

We would like to stress that the generalized FCS gauge theories 
to be constructed in what follows may have applications beyond 
higher spin gravity. 
With this in mind, we shall not make any definite choice for the 
internal associative algebra ${\cal H}\,$, that we shall hence
treat formally, sidestepping temporarily the important issues 
of choices of bases for the star product algebras, related 
function classes and the finiteness of the Lagrangian. 
Our focus is instead on how the nature of the Frobenius algebra 
${\cal F}$ is affected by the gauge invariance and the existence 
of a polarization in target space\footnote{It would be interesting
to also consider polarizations from vector field structures
on the base manifold.}, such that the theory can be 
defined globally on a manifold ${\cal M}$ given by the direct product of 
a commutative manifold with boundaries (that may contains spacetimes), and 
a closed noncommutative manifold.
Eventually, we hope to be able to address the global formulation
on more general noncommutative manifolds (obtained from differential
Poisson structures and their homotopy associative extensions),
though the simplified geometries to be considered here nonetheless 
exhibits enough structure in order to lead to nontrivial constraints
on the underlying Frobenius algebra.

The plan of the paper is as follows: 
In Section~\ref{sec:Section2} we review the original 
FCS model~\cite{Boulanger:2015kfa}. 
In Section~\ref{sec:Section3}, we tackle the problem of 
how this model can be generalized by studying, under 
some assumptions, the most general cubic action for 
a set of odd and even form master fields on an
odd-dimensional noncommutative manifold, which 
leads to the emergence of $\mathbb Z_2$-graded 
quasi-Frobenius algebras.
We then show how the global formulation on direct
product manifolds with boundaries can be achieved
using a polarization in target space, yielding the 
generalized FCS gauge theory action \eqref{ma1};
notably, the attendant inner product need not be
a trace operation as the algebra need not be unital.
In Section~\ref{sec:UnitalAlgebra}, we introduce a unit
element and an outer Kleinian operator such that the 
action can be written as an integral of a Chern--Simons-like 
Lagrangian density expressed using a trace and a single 
odd master field, referred to as the superconnection~\cite{Quillen198589}. 
In Section~\ref{sec:Three-grading}, we provide a general scheme 
for the elimination of all zero-form constraints by employing 
a 3-grading of the Frobenius algebra.
Section~\ref{sec:Examples} contains a set of examples 
based on matrix algebras and twisted group algebras,
containing two new models of interest to higher spin
gravity, namely a $\mathbb Z_2$-graded model with
zero-form constraints containing a self-dual branch,
and an FCS generalization of a set of bosonic Konstein--Vasiliev 
models (with internal symmetries).
We conclude in Section \ref{sec:Conclusions}, pointing
to future directions involving homotopy associative 
algebras and tightening correspondency to underlying 
first-quantized model.
The appendix contains a summary of basic properties
of twisted group algebras, and a demonstration of the
fact that Frobenius algebra of the original FCS model
is a twisting of the $\left({\mathbb Z}_2\right)^3$ 
group algebra.

\section{Review of the minimal FCS gauge theory}
\label{sec:Section2}

The FCS model of \cite{Boulanger:2015kfa} is formulated 
on a direct product manifold
\be {\cal M}_9 = {\cal X}_5 \times {\cal Z}_4\ ,\ee
where ${\cal Z}_4$ is a four-dimensional 
closed noncommutative manifold 
and ${\cal X}_5$ is a five-dimensional open 
commutative manifold whose boundary 
contains spacetime (possibly as an open subset).
The model consists of locally defined differential forms
\footnote{\label{footnote1} The master fields are elements of
$\Omega({\cal X}_{5,\xi})\otimes \Omega(\Comp^4)$,
where ${\cal X}_{5,\xi}$ are coordinate charts of
${\cal X}_5$ and $\Omega(\Comp^4)$ consists of
forms, including distributions, on a real slice of 
$\Comp^4$.
The local representatives are assumed to belong to
sections of a structure group such that the curvatures,
covariant derivatives and Lagrange multipliers that
appear in the Lagrangian obey regularity conditions 
in the interior of $\Comp^4$ and fall-off conditions 
at infinity as to make the action well defined.
The way in which this was achieved in \cite{Boulanger:2015kfa}
provides ${\cal Z}_4$ with the topology of $S^2\times S^2$, 
whereas in general there may exist other possibilities, that
we defer for future work. 
} 
on ${\cal M}_9$, referred to as master fields, 
valued in an associative higher spin algebra
${\cal W}\otimes {\cal K}$, where ${\cal W}$ is 
a Weyl algebra extended by inner Klein operators, 
and ${\cal K}=\Comp(\mathbb Z_2\times \mathbb Z_2)$ 
is the (untwisted) group algebra generated by two 
outer Klein operators of $\Omega({\cal Z}_4)\otimes {\cal W}$; 
for further details, see \cite{Boulanger:2015kfa}.

The spectrum of master fields make up a superconnection $Z$ in 
a generalized bundle space\footnote{The transition
elements of ${\cal E}$ consist of forms in all degrees.
Moreover, in order for the model to consist of bosonic 
fields with integer spins, the space $\Omega({\cal Z}_4)\otimes {\cal W}$,
is projected in a fashion that correlates the dependencies 
on the generating elements of $\Omega(Z_4)$ and
${\cal W}$, that is, the dependences of the sections of 
${\cal E}$ on base and fiber coordinates is intertwined.} 
${\cal E}$ with fiber given by ${\cal A}={{\cal W}\otimes{\cal K}}\otimes {\cal F}$,
where 
\be {\cal F}={\cal F}_0\oplus h{\cal F}_0=
{\cal F}^{(-1)}\oplus {\cal F}^{(0)}\oplus {\cal F}^{(+1)}\ ,\ee
is the associative algebra built from 
\be {\cal F}_0={\rm mat}_2(\Comp)=\bigoplus_{i,j=1,2} \Comp\otimes e_{ij}\ ,\ee
and an outer Klein operator $h$, subject to the product rules
\be e_{ij}e_{kl}=\delta_{jk}e_{il}\ ,\qquad he_{ij}=(-1)^{i-j} e_{ij}h\ ,\qquad h^2=1\ ,\ee
which provides ${\cal F}_0$ with a $\mathbb Z_2$-grading that can be 
further refined into a 3-grading by declaring
\be e_{ij},\, h\, e_{ij}\in {\cal F}^{(j-i)}\ .\ee
The algebra can be realized as
\bea
&e:=e_{11} =
\left(\begin{smallmatrix}
1 & 0 \\ 0 & 0
\end{smallmatrix}\right)\otimes \mathbb{1}\;,\quad
\tilde{e}:=e_{22} =
\left(\begin{smallmatrix}
0 & 0 \\ 0 & 1
\end{smallmatrix}\right)\otimes \mathbb{1}\;,\\
& f:=e_{12} =
\left(\begin{smallmatrix}
0 & 1 \\ 0 & 0
\end{smallmatrix}\right)\otimes \s_1\;,\quad
\tilde{f}:=e_{21} =
\left(\begin{smallmatrix}
0 & 0 \\ 1 & 0
\end{smallmatrix}\right)\otimes \s_1\;,
\\
& h :=
\left(\begin{smallmatrix}
1 & 0 \\ 0 & 1
\end{smallmatrix}\right)\otimes \s_3\;.
\label{44}
\eea
Alternatively, in order to make manifest the 3-grading, 
one may display the algebra as
\be {\cal F}=\left[\begin{array}{cc} e\oplus  he &\ f\oplus hf \\ \tilde f \oplus h\tilde f &\ \tilde e\oplus h \tilde e\end{array}
\right]\ ,\ee
where 
\be {\cal F}^{(-1)}=\tilde f \oplus h\tilde f\ ,\qquad
{\cal F}^{(0)}=e\oplus  he \oplus \tilde e\oplus h \tilde e\ ,\qquad
{\cal F}^{(+1)}=f\oplus hf \ .
\label{2x2}\ee

The superconnection can thus be expanded as
\be
Z=h X+P\ ,
\ee
where
\be
X = \sum _{i,j} X^{ij} e_{ij} = \begin{pmatrix}
A & B \\  \widetilde B & \widetilde A
\end{pmatrix}\ ,\qquad
P = \sum _{i,j} P^{ij} e_{ij} = \begin{pmatrix}
V & U \\ \widetilde U & \widetilde V
\end{pmatrix}\ ,
\label{xp}
\ee
whose entries are ${{\cal W}\otimes{\cal K}}$-valued master fields 
decomposing under the form degree on ${\cal M}_9$ 
as follows\footnote{The restricted spectrum of form degrees 
yields a model without zero-form constraints on-shell and 
zero-form sector identical to that of Vasiliev's original system.
It is possible, however, to take $(B,\widetilde B;U,\widetilde U)$
and $A,\widetilde A;V,\widetilde V)$ to be general even and odd
forms, respectively, as we shall examine in more detail in
Section \ref{sec:Section3}.}
\bea
& {\rm deg} (B,A,\widetilde A,\widetilde B)
\in\left\{(2n,1+2n,1+2n,2+2n)\right\}_{n=0,1,2,3}\ ,
\\
& {\rm deg} (\widetilde U,V,\widetilde V,U)=\left\{(8-2n,7-2n,7-2n,6-2n)
\right\}_{n=0,1,2,3}\ .
\label{contentDE}
\eea
Introducing the $\mathbb Z$-valued superdegree map ${\rm deg}_{{\cal E}}$
given by the sum of form degree and Frobenius degree, \emph{viz.}
\be {\rm deg}_{{\cal E}}:= {\rm deg}_{{\cal M}_9}+{\rm deg}_{\cal F}\ ,\ee
and the superdifferential
\begin{equation}
q:=hd    
\end{equation}
it follows that $Z$, and hence $X$ and $P$, are odd 
elements in strictly positive superdegrees, that is
\be {\rm deg}_{{\cal E}}(Z),\ {\rm deg}_{{\cal E}}(X),\ {\rm deg}_{{\cal E}}(P)\in \{1,3,\dots\}\ ,\ee
and that $q$ is a nilpotent differential of superdegree 
${\rm deg}_{{\cal E}}(q)=1$, \emph{viz.}
\be  q(f\star g) = q(f)\star g + (-1)^{{\rm deg}_{{\cal E}} (f)} f\star q g\ ,\qquad f,g\in{\cal E}\ .\ee

Turning to the action, it requires a (cyclic) trace operation 
\be {\rm Tr}_{{\cal A}}={\rm Tr}_{{\cal W}\otimes{\cal K}}{\rm Tr}_{{\cal F}}\ ,\ee
where ${\rm Tr}_{{{\cal W}\otimes{\cal K}}}={\rm Tr}_{\cal W} {\rm Tr}_{\cal K}$
is composed out of the standard trace on ${\cal K}$ and a modified
supertrace on ${\cal W}$ (making use of the inner Klein operators to 
achieve cyclicity), and 
\be
{\rm Tr}_{{\cal F}}(e_{ij})=\delta_{ij}\;,\quad
{\rm Tr}_{{\cal F}}(h e_{ij})=0\;.
\ee
Letting $\pi_h$ be the automorphism that sends $h$ to $-h$, 
the action proposed in \cite{Boulanger:2015kfa} reads
\be
S =  \int_{{\cal M}_9} {\rm Tr}_{{\cal A}}
\left( \tfrac{1}{2}  \,Z\star q Z+\tfrac{1}{3}\, Z\star Z\star Z\right)
-\tfrac{1}{4}\, \oint_{\partial{\cal M}_9} {\rm Tr}_{{\cal A}} 
\,[h\pi_h(Z) \star Z]\ ,
\label{fcs1}
\ee
or, equivalently,
\be S=
\int_{{\cal M}_9} {\rm Tr}_{{\cal A}} \, \left(P \star F^X
+\tfrac13 \,P\star P\star P\right) \ ,
\label{fcs2}
\ee
where
\be
F^X := dX +hXh\star X\ , 
\label{FX}
\ee
and it is assumed that the locally defined configurations
are glued together (see footnote \ref{footnote1})  
such that the Lagrangian is globally defined and that  
$\int_{{\cal M}_9}{\rm Tr}_{\cal A}$ and $\oint_{\partial{\cal M}_9 }{\rm Tr}_{\cal A}$ are cyclic
and graded cyclic operations, respectively, that are 
non-degenerate and obey Stokes' theorem.

Taking $X$ to fluctuate freely at $\partial{\cal M}_9$, the
variational principle implies\footnote{Beyond the semi-classical 
analysis, the boundary condition on $P$ follows from the 
Batalin-Vilkovisky master equation.}
\be R:=qZ+Z\star Z\approx 0\ ,\qquad  P|_{\partial {\cal M}_9}\approx 0\ ,
\ee
which do not contain any zero-form constraints
since ${\rm deg}_{{\cal E}}(Z\star Z)\geqslant 1$. 
Equivalently, by decomposing $R=R^X+R^P$, where $R^X$ and $hR^P$ are
$h$-independent, the equations of motion can be written as
\be R^X:= F^X+P\star P \approx 0\ ,\qquad R^P:= Q P \approx 0\ ,\ee
with
\be
{Q f  :={q} f + h X\star f -(-1)^{{\rm deg}_{{\cal E}} (f)} f \star hX } \ ,\ee
obeying the graded Leibniz rule
\be
Q(f\star g) 
= Q(f)\star g + (-1)^{{\rm deg}_{{\cal E}} (f)} f\star Q g\ .
\ee
The equations of motion form a Cartan integrable system
with Bianchi identities
\be qR+[Z,R]_\star=0\ ,\ee
or, equivalently,
\be 
Q R^X + [P,R^P]_\star \equiv 0\ ,\qquad QR^P 
- [R^X , P]_\star \equiv 0\ .
\ee
as can be seen using the ordinary Bianchi identities
\be Q^2f=[F^X, f]_\star\ ,\qquad QF^X=0\ .\ee
The generalized Bianchi identities ensure invariance 
of the action under the gauge transformations 
\be
\delta Z= q\theta + [Z,\theta]_\star\ ,
\ee
up to total derivatives.
Decomposing
\be
\theta =\epsilon^X + h \epsilon^P\ ,
\ee
one finds that an $\epsilon^X$-transformation leaves the action invariant, 
while an $\epsilon^P$-transformation yields a total derivative that vanishes 
provided that $\e^P$ belongs to the same section as $P$ and
\be
\e^P|_{\partial{\cal M}_9}=0\ .
\ee

Using the basis \eqref{2x2} to decompose 

\be
Z={\bf A}+{\bf B}\ ,\ee
where 
\be {\bf A}  = \begin{pmatrix}
hA+V & 0 \\ 0 &  h\wt A +\wt V 
\end{pmatrix}\ ,\qquad
{\bf B}  = \begin{pmatrix}
0 & hB + U \\ h\wt B + \wt U  & 0
\end{pmatrix}
\label{AB}
\ee
are odd and even forms, respectively, the action \eq{fcs1} takes the form
\bea
S &=&  \int_{{\cal M}_9} {\rm Tr}_{{\cal A}}
\Big[\tfrac{1}{2}\, {\bf A} \star q {\bf A} +\tfrac{1}{3}\,{\bf A\star \bf A\star \bf A}
+\tfrac12 {\bf B} \star \left( q\bf B + \bf A\star\bf B + \bf B \star \bf A\right)\Big]
\nn\\
&& -\tfrac{1}{4}\, \int_{\partial{\cal M}_9} {\rm Tr}_{{\cal A}} 
\,\Big[h\pi_h({\bf A}) \star {\bf A} + h\pi_h(\bf B) \star \bf B\Big]\\[5pt]
&=&  \int_{{\cal M}_9} {\rm Tr}_{{{\cal W}\otimes{\cal K}}}\Big[ V\star\left( F -  B \star\wt B  
+\tfrac13\, V\star V + U \star\wt U
\right) +
\wt U \star DB  
\nn\\
&&  \ \ \ \ \ \ \ +\wt V\star \left( \wt F - \wt B\star B   
+ \tfrac13\, {\wt V}\star \wt{V}  + \wt U\star U\right) +
U \star \wt D \wt B  \Big]\ ,
\label{SH}
\eea
where we have defined
\bea
& F ={d} A+A\star A\ ,\qquad\qquad\qquad\quad  \widetilde F=
{d} \widetilde A+\widetilde A\star \widetilde A\ ,
\\
& DB = {d} B+A\star B-B\star \widetilde A\ ,\qquad \widetilde D \widetilde B=
{d} \widetilde B+\widetilde A\star \widetilde B-\widetilde B\star A\ .
\eea
Finally, on $\partial{\cal M}_9$, where $(U,\widetilde U;V,\widetilde V)$
vanish, the equations of motion read
\bea
& F - B\star\widetilde B\approx  0\ ,\qquad D B\approx  0\ ,
\\
& \widetilde F-\widetilde B\star B\approx  0\ ,\qquad \widetilde D \widetilde B\approx  0\ ,
\label{E2}
\eea
which can be shown to contain Vasiliev's equations upon 
expanding around a vacuum expectation value 
the dynamical two-form in $\wt B$ and fixing a gauge for
$\wt A-A$; in the simplest setting 
\be  \wt A=A=W\ ,\qquad \widetilde B=J\ ,\ee
where $J$ is a closed a central two-form and the reduced system
takes the form 
\be dW+W\star W+B\star J=0\ ,\qquad dB+W\star B-B\star W=0\ .\ee

Extending the model by $\widetilde B_{[0]}$, ${\widetilde U}_{[0]}$, 
$B_{[8]}$, $U_{[8]}$ and the top-forms $A_{[9]}$, $\wt A_{[9]}$, 
$V_{[9]}$ and $\wt V_{[9]}$, yields a gauge invariant action 
for a superconnection with
\be {\rm deg}_{{\cal E}}(Z),\ {\rm deg}_{{\cal E}}(X),
\ {\rm deg}_{{\cal E}}(P)\in \{-1,1,3,\dots\}\ .\ee
leading to quadratic zero-form constraints on-shell that 
are compatible with the differential constraints 
on the zero-forms\footnote{The zero-form constraints and 
top-forms can be treated within the context of path integral 
quantization using the AKSZ formalism.}; on the boundary
\be B_{[0]}\star \widetilde B_{[0]}\approx 0\ ,\qquad
 \widetilde B_{[0]}\star B_{[0]}\approx 0\ .\ee
The equations of motion can furthermore be 
extended by ten-form curvature constraints for
the top-forms, which yields a universal 
quasi-free differential algebra (from which 
the top-form gauge transforms can be read off).

The algebra ${\cal F}$ reflected in the master field 
content of the model is an example of a \emph{Frobenius algebra}, 
that is, a unital associative algebra with a nondegenerate 
invariant bilinear form\footnote{The positively normed Frobenius 
algebras are $\mathbb{R}$, $\mathbb{C}$ or $\mathbb{H}\,$.}.
In what follows, we shall generalize the above model 
by arranging the master fields using $\mathbb Z_2$-graded but not necessarily 
unital associative algebras with nondegenerate bilinear forms, 
which we shall refer to as $\mathbb Z_2$-graded 
quasi-Frobenius algebras.
Just as in the model above, the $\mathbb Z_2$-grading 
will be crucial for on-shell integrability and gauge invariance,
while the stronger 3-grading, which facilitates the removal of 
top-forms off-shell and hence zero-form constraints on-shell, 
is an optional requirement\footnote{Top-forms and zero-form constraints
can naturally be incorporated into the AKSZ formalism.}.
Although unitality is optional as well, the unital case, which contains
(twisted) group algebras and more general Hopf algebras, is nonetheless 
interesting as it permits the usage of inner Klein operators to generate 
the polarization.

\section{Generalized FCS gauge theory}
\label{sec:Section3}

In this section we shall generalize the FCS model of 
Section 2 to models consisting of a finite numbers of 
master differential forms on odd-dimensional noncommutative 
manifolds with boundaries, valued in an associative algebra
with a trace operation.
Under natural assumptions on the resulting differential 
form algebra, \emph{i.e.} Leibniz' rule, Stokes' law 
and cyclicity of the integration operation, we shall demonstrate 
that the most general cubic covariant Hamiltonian  
action with canonical kinetic term (without terms containing
more than one exterior derivative nor mass terms) is governed by a 
$\mathbb Z_2$-graded quasi Frobenius algebra.

We remark that on a commutative manifold ${\cal M}$, 
the off-shell gauge invariance of a general covariant
Hamiltonian action (with general symplectic potential
and including mass terms) is equivalent to on-shell Cartan 
integrability.
In the noncommutative case, this equivalence continues to hold
in the case of a canonical kinetic term.
Below, in the cubic case, we shall also keep track of boundary terms 
and deal with global formulations by means of specifications of polarizations.

\subsection{Cubic action}
\label{sec:ansatz}

We consider a Lagrangian of the 
form\footnote{The FCS Lagrangian \eq{SH} is obtained 
by taking $A^I=(A,\wt A; V, \wt V)$, $B^P=(B,\wt B;
U ,\wt U)$ and $\left(\Sigma_{IJ},  \Omega_{PQ}, \Theta_{IJ} , 
\Xi_{PQ}\right) = \left(\mathbb{\sigma}_1 \otimes 
\mathbb{1 },-{\rm i}\mathbb{\sigma}_2 \otimes \mathbb{\sigma}_1,
{\rm i}\mathbb{\sigma}_2 \otimes \mathbb{1}, -\mathbb{\sigma}_1
\otimes \mathbb{\sigma}_1\right)$, and making a suitable
identification of $s_{IPQ}$ and $t_{IJK}$.}
\bea
S\!&=&\!\! \int_{\cal M}\!\! {\rm Tr}_{{\cal H}}\! \left( \tfrac{1}{2}\,A^I  \star { d} A^J \,\Sigma_{IJ}
 + \tfrac{1}{2}\,B^P \star { d} B^Q \,\Omega_{PQ} + 
 \tfrac{1}{3}\, {t}_{IJK}A^I\star A^J \star A^K\,
 + s_{IPQ} \, A^I \star B^P \star B^Q \right)
\nonumber \\
 && +  \;\tfrac{1}{4}\oint_{\partial \cal M} {\rm Tr}_{{\cal H}} [ A^I \star A^J \,\Theta_{IJ}
 - B^P \star B^Q \,\Xi_{PQ}]
\label{cubicaction2}
\eea
where ${\cal M}$ is a noncommutative manifold of dimension $2n+1$ with boundary
$\partial{\cal M}$, and 
\begin{eqnarray}
A^I &=& A^I_{[1]}+A^I_{[3]}+\cdots+A^I_{[2n+1]}\;,
\label{AI} \\
B^P &=& B^P_{[0]}+B^P_{[2]}+\cdots+B^P_{[2n]}\;,\label{BP}
\end{eqnarray}
where $I=1,...,N_+$ and $P=1,...,N_-$, which we shall refer to
as the master fields, are differential forms on $\cal M$ valued 
in an associative algebra ${\cal H}$, that is, elements of 
$\Omega({\cal M})\otimes {\cal H}$.
Initially, we shall assume that the master fields are defined globally
on ${\cal M}$; in Section \ref{Sec:3.4} we shall relax this condition
and provide a global formulation in terms of locally defined fields
of a special type on direct product manifolds.

In \eqref{cubicaction2}, the $\star$ denotes the combined associative
product on $\Omega({\cal M})\otimes {\cal H}$ and $\Omega(\partial{\cal M})\otimes {\cal H}$.
It is assumed that the operation of restricting to the boundary
commutes with the star product operation, \emph{i.e.}
$(f\star g)|_{\partial {\cal M}}=(f)|_{\partial\cal M}\star (g)|_{\partial{\cal M}}$
for $f,g\in \Omega({\cal M})\otimes {\cal H}$.
The combined operations $\int_{\cal M} {\rm Tr}_{{\cal H}}$ and 
$\oint_{\cal \partial M} {\rm Tr}_{{\cal H}}$, where ${\rm Tr}_{{\cal H}}$
denotes a trace operation on ${\cal H}$, are assumed to be cyclic and 
graded cyclic linear maps on $\Omega({\cal M})\otimes {\cal H}$
and $\Omega(\partial{\cal M})\otimes {\cal H}$, respectively.
We shall assume that Leibniz' rule holds together with Stokes' 
theorem, \emph{viz.} $\int_{\cal M} {\rm Tr}_{{\cal H}}
d(\cdot)=\oint_{\partial\cal M} {\rm Tr}_{{\cal H}}(\cdot)$.
It follows that 
\bea
\Sigma_{IJ} &=& \Sigma_{JI}\ ,\qquad \Omega_{PQ}= -\Omega_{QP}\ ,\qquad
t_{IJK}= t_{JKI}\ ,
\nn\\
\Theta_{IJ} &=& -\Theta_{JI}\ ,\qquad \Xi_{PQ} = \Xi_{QP}\ .
\label{sym}
\eea
We assume that $\Sigma_{IJ}$ and $\Omega_{PQ}$ are nondegenerate 
with inverses defined by 
\begin{eqnarray}
\Omega^{RP}\Omega_{QP} = \delta^R_Q\ ,\qquad
\Sigma^{IK}\Sigma_{JK} = \delta^I_J\ .
\label{inverse}
\end{eqnarray}
We assume that all master fields are real, \emph{i.e.}
\be (A^I)^\dagger=A^I\ ,\qquad (B^P)^\dagger=B^P\ ,\ee
where the dagger denotes the hermitian conjugation map of 
$\Omega({\cal M})\otimes {\cal H}$,
which is assumed to obey
\be (\int_{\cal M}{\rm Tr}_{{\cal H}} (f\star g))^\dagger=\int_{\cal M}{\rm Tr}_{{\cal H}} (g^\dagger \star f^\dagger)\ ,\qquad f,g\in\Omega(\partial{\cal M})\otimes {\cal H}\ .\ee
Thus, the action is then real provided that
\be (\Sigma_{IJ})^\dagger=\S_{IJ}\ ,\qquad (\Omega_{PQ})^\dagger= \Omega_{PQ}\ ,\ee\be
({t}_{IJK})^\dagger= t_{IKJ}\ ,\qquad (s_{IPQ})^\dagger= s_{IQP}\ ,\qquad
(\Theta_{IJ})^\dagger=-\Theta_{IJ}\ ,\qquad (\Xi_{PQ})^\dagger=\Xi_{PQ}\ .\ee

The total variation 
\begin{eqnarray}
\delta S &=& \int_{\cal M} {\rm Tr}_{{\cal H}} \left[
\delta A^I \star \,R^J\,\Sigma_{IJ} + \delta B^P \star \,R^Q\,\Omega_{PQ} \right]
\nonumber \\
&& +\; \tfrac{1}{2}\,\oint_{\partial{\cal M}} {\rm Tr}_{{\cal H}} \left[
\delta A^I \star \,A^J\,(\Theta_{IJ}+\Sigma_{IJ}) - \delta B^P \star \,B^Q\,
(\Xi_{PQ} + \Omega_{PQ}) \right]\ ,
\label{variaSbulkgeneral}
\end{eqnarray}
where the generalized curvatures
\bea
R^I &:= { d}A^I + {t}^I{}_{JK}\,A^J \star  A^K + s^I{}_{PQ}\,B^P \star B^Q \ ,
\\
R^P &:= { d}B^P - s_{IQ}{}^{P}\,A^I \star B^Q -  s_I{}^P{}_{Q}\, B^Q \star A^I \ ,
\label{eom22}
\eea
and the indices are raised and lowered using the conventions
\be A^I=\Sigma^{IJ}A_J\ ,\qquad B^P=\Omega^{PQ}B_Q\ .\ee
Provided that the coefficients $(t_{IJK},s_{IPQ})$ obey the
quadratic constraints
\bea
&{t}^J{}_{MK}\,{t}^I{}_{LJ} ~= ~ {t}^J{}_{LM}\,{t}^I{}_{JK}\;,
\quad
{s}^I{}_{PQ}\,{s}_{IRT} ~= ~ - {s}^I{}_{TP}\,{s}_{IQR}\;, &
\\
& {s}_{J P}{}^R\, {s}_{I RQ}~= ~  - {t}^K{}_{IJ}\,{s}_{K PQ}\ , \quad
 s_I{}^P{}_Q \, s_{J P R} ~ = ~ s_{J Q}{}^P \, s_{I R P} \;,&
\label{qc}
\eea
the curvatures obey the generalized Bianchi identities
\bea
{ d} R^I - {t}^I{}_{JK} ( R^J \star A^K - A^J \star R^K )
- s^I{}_{PQ} ( R^P \star B^Q + B^P \star R^Q ) & \equiv \; 0\ ,
\\
{ d} R^P + {s}_I{}^P{}_Q ( R^Q \star A^I + B^Q \star R^I )
+ s_{IQ}{}^{P} ( R^I \star B^Q - A^I \star R^Q )  & \equiv\;  0\ .
\eea

In particular, the variation of the action with respect to $A^I_{[2n+1]}$
yield the zero-form constraint 
\be R^I_{[0]}\equiv s^I_{PQ} B^P_{[0]} \star B^Q_{[0]}\approx 0\ .\ee
Its exterior derivative is proportional to $R^P_{[1]}$, that is, 
the bulk equations of motion define a quasi-free differential algebra with 
a zero-form constraint.
Moreover, the curvature $R^I_{[2n+2]}$ of $A^I_{[2n+1]}$ does not 
appear in the variation of the action but can nonetheless be introduced
within the context of a universal quasi-free differential algebra.

The Cartan gauge transformations 
\bea\label{gaugevariaB}
\delta  A^I  &={ d} \epsilon^I
                + {t}^I{}_{JK}(A^J \star \epsilon^K- \epsilon^J \star A^K)
                - s^I{}_{PQ}(\eta^P \star B^Q + B^P\star \eta^Q) \,
\\
\delta B^P  &= { d}\eta^P
                + s_{IQ}{}^{P}(\epsilon^I \star B^Q - A^I\star \eta^Q)
                + s_I{}^P{}_{Q}( \eta^Q \star A^I + B^Q\star \epsilon^I) \ ,
\eea
transform the curvatures covariantly and leave the action invariant
up to boundary terms, which will be studied below in a more 
streamlined notation.

\subsection{$\mathbb Z_2$-graded quasi-Frobenius algebra}
\label{sec:subSection2}

The constraints \eq{sym}, \eq{inverse} and \eq{qc} are equivalent 
the existence of a $\mathbb{Z}_2$-graded associative algebra 
\be {\cal F}={\cal F}^{+}\oplus {\cal F}^{-}\ ,\ee
where  
\be {\cal F}^{+}=\bigoplus_{I=1}^{N^{+}} \Comp \otimes e_I\ ,\qquad
\qquad {\cal F}^{-}=\bigoplus_{P=1}^{N^{-}} \Comp \otimes f_P\ ,\ee
in terms of generators obeying the product laws
\bea
e_I e_J &=  e_K\, t^K{}_{IJ}\ , \qquad f_P f_Q = - e_I \, s^I{}_{PQ}\ ,
\label{fa}\\
e_I f_R &= - f_P s_{I R}{}^P \ , \qquad f_R e_I = f_P\, s_I{}^P{}_R\ ,
\eea
with a non-degenerate bilinear form
\be 
(e_I,e_J)_{\cal F} = \Sigma_{IJ}\ , 
\qquad (f_P,f_Q)_{\cal F} =  \Omega_{PQ} \ ,
\label{bilinear}
\ee
obeying the invariance condition
\be\label{3.22} (a,bc)_{\cal F}=(ab,c)_{\cal F}\ ,\qquad a, b, c\in {\cal F}\ ,\ee
and the graded symmetry property
\be\label{3.23}
(a,b^{\sigma})_{\cal F} = \sigma(b^{\sigma},a)_{\cal F}\ ,
\qquad a\in {\cal F}\ . \quad b^{\sigma} \in {\cal F}^{\sigma}\ ,\qquad \s=\pm\ .
\ee
The associativity conditions $e(ee)=(ee)e, 
f(ff)=(ff)f,e(ef)=(ee)f$ and $e(fe)=(ef)e$ imply the constraints 
in \eq{qc}. 
The invariance conditions
$(e_I e_J,e_K)_{\cal F} = (e_I, e_J e_K)_{\cal F}$ and
$(e_I f_P,f_Q)_{\cal F} = (e_I, f_P f_Q)_{\cal F}$, respectively,
hold by virtue of the cyclicity of $t_{KIJ}$ and the fact
that both $e_I f_P$ and $f_P f_Q$ are given in terms of $s_{IPQ}$.

Introducing the master fields\footnote{
We use the convention that if $f,g \in {{\cal H}}$ 
and $a,b\in {\cal F}$ then $(af,bg)_{\cal F}\equiv (a,b)_{\cal F} f\star g $.}
\begin{eqnarray}
\boldsymbol{A} :=\sum_I A^I e_I\in {{\cal H}}\otimes {\cal F}^{+}\; ,\qquad
\boldsymbol{B} := \sum_P B^P f_P\in {{\cal H}}\otimes {\cal F}^{-}\; ,
\label{mf1}
\end{eqnarray}
and corresponding curvatures
\be 
\boldsymbol{F}= d\boldsymbol{A}+\boldsymbol{A}\star \boldsymbol{A}\ ,
\qquad \boldsymbol{D}\boldsymbol{B}=d\boldsymbol{B}+\boldsymbol{A}\star \boldsymbol{B}-
\boldsymbol{B}\star\boldsymbol{A}\ ,
\ee
the action (\ref{cubicaction2}) can be re-written as
\bea
S &=& \int_{\cal M} {\rm Tr}_{{\cal H}} \left[ \tfrac{1}{2}
(\boldsymbol{A},d\boldsymbol{A})_{\cal F} + \tfrac13 (\boldsymbol{A},
\boldsymbol{A}\star \boldsymbol{A})_{\cal F}
+\tfrac{1}{2}(\boldsymbol{B},\boldsymbol{D}\boldsymbol{B})_{\cal F}\right]
\nn\w2
&& -  \;\tfrac{1}{4}\oint_{\partial \cal M} {\rm Tr}_{{\cal H}} \left[(\boldsymbol{A},
\Theta(\boldsymbol{A}))_{\cal F} -(\boldsymbol{B},\Xi (\boldsymbol{B}))_{\cal F}\right]\ ,
\label{ba}
\eea
where we have defined the outer operators
\be \Theta (e_I):=\Theta_I{}^J e_J\ ,\qquad \Xi(f_P):=\Xi_P{}^Q f_Q\ .\ee
Given the symmetry properties of $\Theta_{IJ}$ and $\Xi_{PQ}$, these operators obey
\bea 
(a^{+},\Theta (b^{+}))_{\cal F} &= -(b^{+},\Theta (a^{+}))_{\cal F}=-(\Theta (a^{+}),b^{+})_{\cal F}\ ,
\\
(a^{-}, \Xi (b^{-}))_{\cal F} &= (b^{-},\Xi (a^{-}))_{\cal F}=-(\Xi (a^{-}),b^{-})_{\cal F}\ ,
\label{ab}
\eea
for $a^{\pm},b^{\pm}\in {\cal F}^{\pm}$.

Using the above notation, the general variation reads
\begin{eqnarray}
\delta S &=& \int_{\cal M} {\rm Tr}_{{\cal H}} \left[
(\delta\boldsymbol{A}, \boldsymbol{R}^A)_{\cal F} +
(\delta \boldsymbol{B}, \boldsymbol{R}^B)_{\cal F} \right]
\nonumber \\
&& +\; \tfrac{1}{2}\,\oint_{\partial{\cal M}} {\rm Tr}_{{\cal H}} \left[
(\delta \boldsymbol{A},(1-\Theta) \boldsymbol{A})_{\cal F} -
( \delta\boldsymbol{B}, (1-\Xi)\boldsymbol{B})_{\cal F} \right]\ ,
\label{deltaS}
\end{eqnarray}
where the generalized curvatures
\be
\boldsymbol{R}^A := \boldsymbol{F} - 
\boldsymbol{B}\star \boldsymbol{B}\ ,\qquad
\boldsymbol{R}^B := \boldsymbol{D}\boldsymbol{B}\ ,
\label{boldfaceR}
\ee
obey the generalized Bianchi identities
\be \boldsymbol{D} \boldsymbol{R}^A + \{ \boldsymbol{B}, \boldsymbol{R}^B\}_\star \equiv0
\ ,\qquad  \boldsymbol{D} \boldsymbol{R}^B+[ \boldsymbol{B}, \boldsymbol{R}^A]_\star \equiv0\ .
\ee
The Cartan gauge transformations are given by
\be
\delta \boldsymbol{A}=\boldsymbol{D}\boldsymbol{\e}
+\{\boldsymbol{\eta},\boldsymbol{B}\}_\star
\ ,\qquad \delta \boldsymbol{B}=\boldsymbol{D}\boldsymbol{\eta}
-[\boldsymbol{\e},\boldsymbol{B}]_\star\ .
\ee
Under these transformations, the bulk term in \eq{deltaS} 
becomes a total derivative, \emph{viz.}
\be
\int_{\cal M} {\rm Tr}_{{\cal H}} \left[
(\delta\boldsymbol{A}, \boldsymbol{R}^A)_{\cal F} +
(\delta \boldsymbol{B}, \boldsymbol{R}^B)_{\cal F} \right] = 
\oint_{\partial\cal M} {\rm Tr}_{{\cal H}} \left[
(\boldsymbol{\epsilon}, \boldsymbol{R}^A)_{\cal F} +
( \boldsymbol{\eta}, \boldsymbol{R}^B)_{\cal F} \right]\ .
\ee
Taking into account the remaining boundary term in \eq{deltaS} and defining 
\be 
\boldsymbol{A}_\pm=\tfrac12 (1\pm \Theta) \boldsymbol{A}\ ,
\qquad 
\boldsymbol{B}_\pm= \tfrac12 (1\pm \Xi) \boldsymbol{B}\ ,
\label{pros}
\ee
the gauge variation of the action can be written as
\bea 
\delta_{\boldsymbol{\epsilon,\eta}} S &=& \oint_{\partial M} {\rm Tr}_{{\cal H}} \left[
\Big(\boldsymbol{\epsilon}, d \boldsymbol{A}_++
\boldsymbol{A}_+\star\boldsymbol{A}_+
-\boldsymbol{A}_-\star  \boldsymbol{A}_-
-\boldsymbol{B}_+\star \boldsymbol{B}_+
+\boldsymbol{B}_-\star \boldsymbol{B}_-
\Big)_{\cal F}\right.\nn\w2
&&\left. +\Big(\boldsymbol{\eta}, d \boldsymbol{B}_++[\boldsymbol{A}_+,\boldsymbol{B}_+]_\star
+[\boldsymbol{B}_-,\boldsymbol{A}_-]_\star\Big)_{\cal F}\right]\ .
\label{dgs2}
\eea

As we shall see next, the expressions for the variations of the action
given in \eq{deltaS} and \eq{dgs2}, respectively, facilitates the 
global formulation of the model on topologically sufficiently simple 
base manifolds.

\subsection{Polarization in target space}

In what follows we shall give a set of conditions on
$\Theta$, $\Xi$ and the structure coefficients of ${\cal F}$ 
such that the boundary terms in the variations \eq{deltaS} 
and \eq{dgs2} of the action can be expressed in terms of 
representations of a generalized structure group 
(whose transition elements are sums over forms of different 
degrees).

To this end, we begin by observing that since 
$\delta{\boldsymbol{A}}$ and $\delta{\boldsymbol{B}}$ 
are sections, it follows from the form of the total
variation \eq{deltaS} that $(\boldsymbol{A}_-,\boldsymbol{B}_-)$ 
and hence $(\boldsymbol{\e}_-,\boldsymbol{\eta}_-)$ must be sections as well.
Thus, the maximal possible structure group is gauged by 
$\boldsymbol{A}_+$ and $\boldsymbol{B}_+\,$.

Turning to the gauge variation \eq{dgs2}, requiring it
to be writable in terms of sections leads to constraints 
on the structure constants, 
the inner product, $\Theta_{IJ}$ and $\Xi_{PQ}\,$, which 
we refer to as the polarization conditions.
To exhibit these, we assume that $\tfrac12 (1\pm \Theta) $ 
and $\tfrac12 (1\pm \Xi)$ are projectors, that is
\be 
\Theta^2={\rm Id}_{{\cal F}^{+}}\ ,\qquad \Xi^2={\rm Id}_{{\cal F}^{-}}\ ,\label{pc1}
\ee
so that we can decompose
\be {\cal F}_\pm^{+}:=\tfrac12 (1\pm \Theta) {\cal F}^{+}\ ,\qquad 
{\cal F}_\pm^{-}:=\tfrac12 (1\pm \Xi) {\cal F}^{-}\ ,\ee
where thus 
\be ({\cal F}_\pm^{\sigma},{\cal F}_\pm^{\sigma})_{\cal F}=0\ ,\qquad \s=\pm\ ,\ee
in view of \eq{ab}.
Thus, requiring the gauge variation \eq{dgs2} to be expressible in terms of 
sections yields 
\be
({\cal F}_\pm^{+})^{\star 2}\subseteq {\cal F}_+^{+}\ ,\qquad
({\cal F}_\pm^{-})^{\star 2}\subseteq {\cal F}_+^{+}\ ,
\label{mc}
\ee
\be
{\cal F}_{\pm}^{\sigma}\star {\cal F}_{\pm}^{-\sigma}\subseteq {\cal F}_+^{-}
\ ,\qquad \s=\pm\ ,
\label{mc2}
\ee
which are linear constraints on the structure constants $(t_{IJK}, s_{IPQ})$
that together with Eq. \eq{pc1} form the aforementioned polarization conditions.

In order to exhibit the resulting structure, we define
\be 
({\cal A},{\cal B};\overline {\cal U},{\overline {\cal V}}):=(\boldsymbol{A}_+,\boldsymbol{B}_+;
\boldsymbol{B}_-,\boldsymbol{A}_-)\ ,
\label{mfd}
\ee
\be 
(\e^{\cal A},\e^{\cal B};\eta^{\overline {\cal U}},\eta^{\overline {\cal V}}):= ({\boldsymbol\e}_+,{\boldsymbol\eta}_+;
{\boldsymbol\eta}_-,{\boldsymbol\e}_-)\ ,
\label{mpd}
\ee
where thus ${\overline {\cal U}}$ and ${\overline {\cal V}}$ and their gauge parameters belong to sections.
Defining 
\be 
{\cal F}=d{\cal A}+{\cal A}\star {\cal A}\ ,\qquad {\cal D}{\cal B}=d{\cal B}+[{\cal A},{\cal B}]_\star\ ,
\label{fdb}
\ee
and combining \eq{mc} and \eq{mc2} with the fact that the 
only nonvanishing inner products are $(a^\sigma_{\pm},b^\sigma_{\mp})_{\cal F}$, 
from \eq{ba} we arrive at the following action:
\be\boxed{
S=\int_{\cal M} {\rm Tr}_{{\cal H}}\Big[ \left({\overline {\cal U}},{\cal D}{\cal B}\right)_{\cal F}
+\left({\overline {\cal V}},{\cal F} - {\cal B}\star {\cal B}+\tfrac13 {\overline {\cal V}}\star {\overline {\cal V}} 
-{\overline {\cal U}}\star {\overline {\cal U}}\right)_{\cal F}\Big]}
\label{ma1}
\ee
which underlies the general Frobenius--Chern--Simons model based on 
a $\mathbb Z_2$-graded quasi-Frobenius algebra. In contrast to the case of minimal FCS model 
where a trace operation in the Frobenius algebra arises, here the inner product occurs. In summary, 
the route from the general Ansatz in \eq{cubicaction2} to the action \eq{ma1} 
makes use of the equations \eq{fa}, \eq{bilinear}, \eq{mf1},\eq{ba}, \eq{pros} and \eq{mfd}.
If there are no even forms, the action is given by the difference of two 
generalized CS actions on ${\cal M}$ for odd forms ${\cal A}_L$ and ${\cal A}_R$ 
valued in ${\cal H}\otimes {\cal F}^{(+)}$ and with ${\cal A}={\cal A}_L+{\cal A}_R$ 
and $\overline{\cal V}={\cal A}_L-{\cal A}_R$. 

The above action is of the covariant Hamiltonian form, that is, the Lagrange multipliers 
$({\overline {\cal U}},{\overline {\cal V}})$ and the fields $({\cal B},{\cal A})$ belong to dual spaces, since 
the nondegeneracy of the inner product together with Eq. \eq{pc1} imply that 
\be 
{\rm dim}\, {\cal F}_+^{\sigma}= {\rm dim}\,{\cal F}_-^{\sigma}=\tfrac12 N^{\sigma}\ ,
\qquad \s=\pm\ .
\label{dimequal}
\ee
The total variation of \eq{ma1}, which can also be obtained 
from \eq{deltaS}, reads
\bea
\delta S &=&\int_{\cal M} {\rm Tr}_{{\cal H}}\Big[ (\delta {\overline {\cal U}},R^{\cal B})_{\cal F}
+(\delta {\cal B},R^{\overline {\cal U}})_{\cal F}+(\delta {\overline {\cal V}},R^{\cal A})_{\cal F}+
(\delta {\cal A}, R^{\overline {\cal V}})_{\cal F}\Big]
\nn\\
&& +\oint_{\partial{\cal M}}{\rm Tr}_{{\cal H}}
\Big[(\delta {\cal A},{\overline {\cal V}})_{\cal F} -(\delta {\cal B},{\overline {\cal U}})_{\cal F}\Big]\ ,
\label{genv}
\eea
where the Cartan curvatures
\be R^{\cal A} :={\cal F} - {\cal B}\star {\cal B}+{\overline {\cal V}}\star {\overline {\cal V}} - {\overline {\cal U}}\star {\overline {\cal U}}\ ,\qquad R^{\cal B} :={\cal D}{\cal B} - [{\overline {\cal U}},{\overline {\cal V}}]_\star\ ,\ee
\be R^{\overline {\cal V}} :={\cal D}{\overline {\cal V}}-\{{\overline {\cal U}},{\cal B}\}_\star\ ,\qquad R^{\overline {\cal U}} :={\cal D}{\overline {\cal U}}+ [{\overline {\cal V}},{\cal B}]_\star \ ,\ee
obey the generalized Bianchi identities 
\bea
&{\cal D}R^{\cal A}+\{{\cal B},R^{\cal B}\}_\star+[{\overline {\cal V}},R^{\overline {\cal V}}]_\star +\{{\overline {\cal U}},R^{\overline {\cal U}}\}_\star\equiv0\ ,
\w2
&{\cal D}R^{\cal B}+[{\cal B},R^{\cal A}]_\star+[{\overline {\cal U}},R^{\overline {\cal V}}]_\star+\{{\overline {\cal V}},R^{\overline {\cal U}}\}_\star\equiv0\ ,
\w2
&{\cal D}R^{\overline {\cal V}}+[{\overline {\cal V}},R^{\cal A}]_\star+\{{\overline {\cal U}},R^{\cal B}\}_\star+\{{\cal B},R^{\overline {\cal U}}\}_\star\equiv0\ ,
\w2
&{\cal D}R^{\overline {\cal U}}+[{\overline {\cal U}},R^{\cal A}]_\star+\{{\overline {\cal V}},R^{\cal B}\}_\star +[{\cal B},R^{\overline {\cal V}}]_\star\equiv0\ .
\eea
The gauge transformations take the form  
\bea
&\delta {\cal A} ={\cal D}\e^{\cal A}+\{{\cal B},\e^{\cal B}\}_\star+[{\overline {\cal V}},\eta^{\overline {\cal V}}]_\star
+\{{\overline {\cal U}},\eta^{\overline {\cal U}}\}_\star\ ,
\w2
&\delta {\cal B} ={\cal D}\e^{\cal B}+[{\cal B},\e^{\cal A}]_\star+[{\overline {\cal U}},\eta^{\overline {\cal V}}]_\star+\{{\overline {\cal V}},\eta^{\overline {\cal U}}\}_\star\ ,
\w2
&\delta {\overline {\cal V}} ={\cal D}\eta^{\overline {\cal V}}+[{\overline {\cal V}},\e^{\cal A}]_\star+ \{{\cal B},\eta^{\overline {\cal U}}\}_\star+\{{\overline {\cal U}},\e^{\cal B}\}_\star\ ,
\w2
&\delta {\overline {\cal U}} ={\cal D}\eta^{\overline {\cal U}}+[{\overline {\cal U}},\e^{\cal A}]_\star +\{{\overline {\cal V}},\e^{\cal B}\}_\star+[{\cal B},\eta^{\overline {\cal V}}]_\star\ .
\eea
The gauge variation of the action is given by
\be 
\delta_{\boldsymbol{\e,\eta}} S =\oint_{\partial{\cal M}}{\rm Tr}_{{\cal H}}
\Big[\left(\eta^{\overline {\cal U}} , {\cal D}{\cal B}
+[{\overline {\cal U}},{\overline {\cal V}}]_\star\right)_{\cal F}
+\left(\eta^{\overline {\cal V}},{\cal F} - {\cal B}\star {\cal B} 
-{\overline {\cal V}}\star {\overline {\cal V}}
+{\overline {\cal U}}\star {\overline {\cal U}}\right)_{\cal F}\Big]\ .
\label{dgs}
\ee
This result, as well as the result of general variation formula \eq{genv} will 
be used below in studying the global formulation.

\subsection{Global formulation}\label{Sec:3.4}

In order to treat master fields that are defined locally
we need to assume that the integration measure on ${\cal M}$
provides a cyclic trace operation on the algebra of 
locally defined forms.

The polarization introduced above suffices for nontrivial
global formulations on direct product manifolds 
\be {\cal M}={\cal X}\times {\cal Z}\ ,\ee
where ${\cal X}$ is a commuting manifold consisting 
of charts ${\cal X}_\xi$ and ${\cal Z}$ is a closed 
noncommutative manifold for which $\int_{\cal Z}$ 
provides a trace operation on $\Omega({\cal Z})$,
that is cyclic and graded-cyclic, respectively, 
in case ${\rm dim}({\cal Z})$ is odd and even.
The master fields are taken to be locally defined
forms on $\Omega({\cal X}_\xi\times {\cal Z})$.
%
The locally defined configurations can be glued together 
into sections of a generalized bundle\footnote{In the case
of a cubic action being considered here, 
the gluing compatibility condition for a generalized bundle
holds identically for any choice of structure group, 
see \cite{Boulanger:2011dd,Sezgin:2011hq,Boulanger:2012bj}.}, 
that we shall denote by
${\cal E}$, using transition  functions $T_\xi^\eta$ generated by parameters
\be 
((t^{\cal A})_\xi^\eta, (t^{\cal B})_\xi^\eta)\in 
\Omega({\cal X}_\xi\cap {\cal X}_\eta)\times {\cal Z})\ ,
\ee
valued in subspaces of the spaces of ${{\cal H}}\otimes {\cal F}$
that contain the parameters $(\e^{\cal A},\e^{\cal B})$.
Letting ${\cal X}'_\xi\subseteq {\cal X}_\xi$ be patches such that\footnote{
Instead of working with patches one may use partitions of unity.}
\be {\cal X}=\cup_\xi {\cal X}'_\xi\ ,\ee
we may write
\be S=\sum_\xi \int_{{\cal X}'_\xi} \check {\cal L}_\xi 
 \ ,\ee
where the locally defined Lagrangian
\be \check {\cal L}_\xi = \oint_{\cal Z} {\rm Tr}_{{\cal H}} 
\Big[ \left({\overline {\cal U}},{\cal D}{\cal B}\right)_{\cal F}
+\left({\overline {\cal V}},{\cal F} - {\cal B}\star {\cal B}
+\tfrac13 {\overline {\cal V}}\star {\overline {\cal V}} 
-{\overline {\cal U}}\star {\overline {\cal U}}\right)_{\cal F}\Big]_\xi\ .
\ee
Since \eq{dgs} does not contain the parameters $(\e^{\cal A},\e^{\cal B})$, it follows that 
$\check {\cal L}_\xi$ remains invariant (pointwise on ${\cal X}_\xi$)
as the fields are transformed by transition functions.
Thus 
\be \check {\cal L}_\xi=\check {\cal L}|_{{\cal X}_\xi}\ ,\ee
where $\check {\cal L}$ is a globally defined top form on ${\cal X}$,
which is to say that the action is globally defined modulo boundary terms.
The total variation of the action on-shell as well as its gauge variation 
are thus given by terms evaluated at the boundary 
\be \partial{\cal M}=\partial{\cal X}\times {\cal Z}\ ,\ee
that vanish provided that 
\be ({\overline {\cal U}},{\overline {\cal V}}; \eta^{\overline {\cal U}}, 
\eta^{\overline {\cal V}})\rvert_{\partial{\cal M}}=0\ ,\ee
thus leading to a globally defined action including boundary terms.

\section{Unital algebras with Klein operators}
\label{sec:UnitalAlgebra}

In this section, we assume that ${\cal F}$ 
contains a unity, which implies that the inner 
product on ${\cal F}$ is a supertrace.
We also assume that the polarization is 
achieved 
by adding an outer Klein operator $h$ to a 
Frobenius subalgebra ${\cal F}_0\,$ that
is $\mathbb Z_2$-graded with respect to it. 
The resulting FCS model can be formulated succinctly in terms 
of a single master field $Z\in {{\cal H}}\otimes {\cal F}$,
referred to as the superconnection, allowing the inclusion
of higher powers of fields into the action.
%

\subsection{Trace operation and outer Klein operator}

In what follows, we shall assume ${\cal F}$ to be unital,
which implies that the inner product is equivalent to the
nondegenerate graded cyclic supertrace operation
\be 
{\rm STr}_{\cal F}(a):= (1,a)_{\cal F}\ ,\qquad a\in{\cal F}\ ,
\ee
whose graded cyclicity follows from the fact that
\be 
{\rm STr}_{\cal F}(ab^\pm)=(1,ab^{\pm})_{\cal F}=(a,b^{\pm})_{\cal F}
=\pm (b^{\pm},a)_{\cal F}=  \pm\, {\rm STr}_{\cal F}(b^{\pm} a)\ ,\
\ee
for all $a\in {\cal F}$ and $b^{\pm}\in {\cal F}^{\pm}$.

We furthermore assume that ${\cal F}$ contains an idempotent 
element $h$, referred to as the Klein operator of the 
$\mathbb Z_2$-graded algebra, such that
\bea
h a^{\pm} h=\pm a^{\pm}\ ,\qquad h^2=1\ ,\qquad
a^{\pm}\in {\cal F}^{\pm}\ .
\eea
Inserting this operator into the supertrace yields the
nondegenerate (cyclic) trace operation
\be 
{\rm Tr}_{\cal F}(a):= {\rm STr} (ha) \ ,\qquad a\in{\cal F}\ .
\label{trace}\ee
In view of \eq{dimequal}, the polarization conditions 
\eq{mc} and \eq{mc2}, which ensure that 
\be {\cal F}_0:= {\cal F}^{+}_+\oplus {\cal F}^{-}_+\ ,\label{calF0}\ee
is an associative subalgebra of ${\cal F}$, can be solved by 
taking 
\be {\cal F}={\cal F}_0\oplus h{\cal F}_0\ ,
\qquad {\cal F}^{+}_-= h {\cal F}^{+}_+\ ,\qquad 
{\cal F}^{-}_-= h {\cal F}^{-}_+\ ,\label{defh}\ee
that is, by taking $h$ to be outer with respect to ${\cal F}_0$, 
and requiring that ${\cal F}_0$ equipped with the inner product
\be (a,b)_{{\cal F}_0} := (a,b)_{\cal F}\ ,\qquad a,b \in {\cal F}_0\ ,\ee
or, equivalently, the trace operation 
\be {\rm Tr}_{{\cal F}_0}(ab):= {\rm Tr}_{\cal F}(ab)\ ,\qquad a,b \in {\cal F}_0\ ,\ee
is a Frobenius algebra.
In other words, we assume that $1\in {\cal F}_0$ and that 
$ (\cdot,\cdot)_{{\cal F}_0}$ is nondegenerate, after which 
we can define the element $h$ via \eq{defh}.
%

\subsection{Superconnection}

In view of \eqref{calF0} and \eqref{defh}, we introduce the
superconnection
\be
Z = hX+P\ ,\qquad 
X ={\cal A}+{\cal B}\ ,\qquad  P=h({\overline {\cal U}}+{\overline {\cal V}})\ ,
\label{zxp1}
\ee
where thus both $X,P\in {\cal F}_0$, and the superdifferential
\be q=hd\ .\ee
Thus, by letting $\pi_h$ denote the automorphism of ${\cal F}$ 
that sends $h$ to $-h$ while acting as the identity on ${\cal F}_0$, 
the action \eq{ma1} takes the compact form
\be\boxed{
\begin{array}{lcl}S &=&  \int_{{\cal M}} {\rm Tr}_{\cal H\otimes\cal F}
\left( \tfrac{1}{2}  \,Z\star q Z+\tfrac{1}{3}\, Z\star Z\star Z\right)
-\frac{1}{4}\, \oint_{\partial{\cal M}} {\rm Tr}_{{{\cal H}}\otimes{\cal F}} 
\,[h\pi_h(Z) \star Z]\ 
\w2
&=& \int_{\cal M} {\rm Tr}_{{{\cal H}}\otimes{\cal F}_0} \, \left(P \star F^X
+\tfrac13 \,P\star P\star P\right) \ , \end{array}}
\label{n2}
\ee
where
\be F^X=dX+hXh\star X\ .\ee
As for the global definition of the theory, we recall 
that the structure group is generated by a subalgebra 
of the algebra gauged by $X$, and that 
\be P|_{\partial{\cal M}}=0\ .\ee


\subsection{Component formulation}


Using \eq{zxp1} and defining
\be {\cal U}=h{\overline {\cal U}}\ ,\qquad {\cal V}=h{\overline {\cal V}}\ ,\ee
such that $P={\cal U}+{\cal V}$, the action \eq{ma1} can be written as
\be
\boxed{ S=  \int_{\cal M} {\rm Tr}_{{{\cal H}} 
\otimes {\cal F}_0} \Big[ {\cal U} \star {\cal D}{\cal B}
+ {\cal V}\star\left({\cal F} - {\cal B}\star {\cal B}
+ {\cal U}\star {\cal U}+\tfrac13 {\cal V}\star {\cal V} \right) \Big]}\ ,
\label{n3}
\ee
with ${\cal F}$ and ${\cal D B}$ from \eq{fdb}. The form of this action resembles that of 
the action \eq{SH} for the 4D FCS higher spin gravity model reviewed in Section 2, 
though the Frobenius algebra and the attendant trace operation used in \eq{n3} is general.
The action can be given explicitly by splitting
\be
e_I =(e_i,e^i)\ ,\qquad f_P =(f_p,f^p)\ ,\qquad
e^i=he_i= e_i h\ ,\qquad f^p=hf_p=-f_p h\ , 
\label{splitplit}
\ee
where $(e_i,f_p)$ is a basis for ${\cal F}_0$ with
product rules
\be e_i e_j = e_k \, t^k{}_{ij}\;, \qquad
f_p f_q = - e_i\, s^i{}_{pq}\;,\qquad e_i f_p = - s_{ip}{}^{q}\,f_q\;,\qquad
f_q e_i = f_p\, s_i{}^p{}_q\;,
\ee
subject to associativity conditions given by \eq{qc} with majuscule indices
replaced by minuscule indices.
Thus
\be
{\cal F}^{+}_+=\bigoplus_{i=1}^{\tfrac12 N^{+}} \Comp \otimes e_i\ ,\qquad
{\cal F}^{-}_+=\bigoplus_{p=1}^{\tfrac12 N^{-}} \Comp \otimes f_p\ ,\ee
\be
{\cal F}^{+}_-=\bigoplus_{i=1}^{\tfrac12 N^{+}} \Comp \otimes e^i\ ,\qquad
{\cal F}^{-}_-=\bigoplus_{p=1}^{\tfrac12 N^{-}} \Comp \otimes f^p\ ,\ee
and the fields can be expanded as 
\be
{\cal A} = \sum_i A^i e_i\ ,\qquad  {\cal B}= \sum_p B^p f_p \ ,\ee
\be {\cal V}=\sum_i V^i e_i\ , \qquad  {\cal U}= \sum_p 
U^p f_p\ .\ee
The inner product matrices are taken to be
\be
\Sigma_{IJ}=\left[\ba{cc} 0& \delta_i^j\\\delta_j^i&0\ea\right]\ ,\quad
\Omega_{PQ}=\left[\ba{cc} 0& -\delta_p^q\\\delta_q^p&0\ea\right]\ ,
\label{SigmaandOmega}
\ee
such that 
\be \Theta_{IJ}=\left[\ba{cc} 0& \delta_i^j\\-\delta_j^i&0\ea\right]\ ,\quad
\Xi_{PQ}=\left[\ba{cc} 0& -\delta_p^q\\-\delta_q^p&0\ea\right]\ .\ee

In summary so far, starting from the Ansatz \eqref{cubicaction2} for a gauge invariant action, 
including boundary terms, and assuming that the resulting $\mathbb Z_2$-graded 
quasi-Frobenius algebra ${\cal F}$ (as in Section 3.2) in addition
\begin{itemize} \item[i)] obeys the polarization conditions \eqref{mc} and \eqref{mc2} 
under the assumption that Eq. \eq{pc1} holds;
\item[ii)] contains a unity (as in Section 4.1); and
\item[iii)] is $\mathbb Z_2$-graded by means of a Klein operator  
$h\in{\cal F} $ leading to the decomposition \eq{defh}; 
\end{itemize}
we arrive at the action \eqref{n3} with master fields in 
${{\cal H}}\otimes{\cal F}_0$, where ${\cal F}_0$ is the proper 
Frobenius subalgebra of ${\cal F}$ defined in \eq{calF0}.


\section{3-grading}\label{sec:Three-grading}


In this section we shall consider models in which the 
$\mathbb Z_2$-grading is extended into a 3-grading that 
allows the truncation of top-forms off-shell to achieve 
equations of motion that do not contain any algebraic 
zero-form constraints.
We shall then describe a general scheme to obtain the 
3-grading by assuming that the $\mathbb Z_2$-grading of ${\cal F}_0$ 
is achieved by an inner Klein operator $\gamma\in {\cal F}_0$.

\subsection{On-shell free differential algebra}

As shown in Section 3, the $\mathbb Z_2$-grading suffices for 
constructing globally defined actions including top-forms leading 
to equations of motion with zero-form constraints.
The system can be constrained algebraically off-shell
as to remove the top-forms and hence the zero-form constraints 
on-shell, provided that the algebra admits a three grading defined by
\be {\cal F}^{(0)}:={\cal F}^{+}\ ,\qquad {\cal F}^{(-1)}\oplus {\cal F}^{(+1)}:={\cal F}^{-}
\ ,\ee
and ${\cal F}^{(k)}\equiv 0$ for $k=\pm 2, \pm3, \dots$, such that
\be {\cal F}^{(k)}\star {\cal F}^{(k')}\subseteq {\cal F}^{(k+k')}\ .\label{3grade2}\ee
Defining the $\mathbb Z$-valued superdegree map 
\be {\rm deg}_{{\cal E}}:= {\rm deg}_{{\cal M}}+{\rm deg}_{\cal F}\ ,\ee
all top-forms as well as a subset of the next-to-top and zero-forms 
can be set to zero off-shell by imposing
\be {\rm deg}_{{\cal E}}(\boldsymbol{A})\ ,{\rm deg}_{{\cal E}}(\boldsymbol{B}) 
\in \{1,3,\dots,2n-1\}\ .\ee
It follows that the curvatures in \eqref{boldfaceR} obey
\be {\rm deg}_{{\cal E}}(\boldsymbol{R}^A),\ {\rm deg}_{{\cal E}}(\boldsymbol{R}^B)
\in \{2,4,\dots,2n\}\ .\ee
Thus, the truncation is consistent with the equations of motion,
and leads to a free differential algebra on-shell, since
\be \boldsymbol{R}^B_{[0]}=-\boldsymbol{B}_{[0]}\star \boldsymbol{B}_{[0]}=
-\boldsymbol{B}^{(+1)}_{[0]}\star \boldsymbol{B}^{(+1)}_{[0]}\equiv 0\ ,\ee
by \eqref{3grade2}.
Hence, since the algebra is free universally, it follows by a general
lemma that the action is gauge invariant.

\subsection{3-grading from inner Klein operator of ${\cal F}_0$}
\label{Sec:5.2}

Let us assume that ${\cal F}_0$, which is a unital Frobenius algebra
by the assumptions made so far, contains an inner Klein operator 
$\gamma$ that is compatible with $h$ in the sense that 
\be [h,\gamma]=0\ ,\qquad \gamma a^\pm =\pm a^\pm\gamma\quad\mbox{for all}\quad 
a^\pm \in {\cal F}_0^{\sigma}\ ,\qquad \gamma^2=1\ .
\label{gammagrading}\ee
We can then introduce the following $3$-grading
\be {\cal F}=\bigoplus_{q=\pm 1,0} {\cal F}^{(q)}\ ,\qquad
{\cal F}^{(\pm 1)}= \frac12(1\pm \gamma) {\cal F}^{-}\ ,\qquad 
{\cal F}^{(0)}={\cal F}^{+}\ ,\ee
and decompose
\be {\cal F}^{(0)}={\cal F}^{(-0)}\oplus {\cal F}^{(+0)}\ ,\qquad
{\cal F}^{(\pm 0)}=\frac12(1\pm \gamma) {\cal F}^{+}\ ,\ee
such that  
\be \label{4.28}{\cal F}^{(\s 0)}={\cal F}^{(\s 1)}{\cal F}^{(-\s 1)}\ ,\qquad \s=\pm.\ee
Thus, in effect, ${\cal F}$ has the following two by two block structure:
\be {\cal F}=\left[\begin{array}{cc} {\cal F}^{(+0)}& {\cal F}^{(+1)}\\
{\cal F}^{(-1)}& {\cal F}^{(-0)}\end{array}\right]=
\left[\begin{array}{cc} \frac12(1+ \gamma){\cal F}^{+}& \frac12(1+ \gamma){\cal F}^{-}\\
\frac12(1- \gamma){\cal F}^{-}& \frac12(1- \gamma){\cal F}^{+}\end{array}\right]\ .\ee
In particular, 
\be {\cal F}_0=\left[\begin{array}{cc} {\cal F}^{(+0)}_0& {\cal F}^{(+1)}_0\\
{\cal F}^{(-1)}_0& {\cal F}^{(-0)}_0\end{array}\right]=
\left[\begin{array}{cc} \frac12(1+ \gamma){\cal F}^{+}_0& \frac12(1+ \gamma){\cal F}^{-}_0\\
\frac12(1- \gamma){\cal F}^{-}_0& \frac12(1- \gamma){\cal F}^{+}_0\end{array}\right]\ .\ee
Thus, upon expanding 
\be X=\left[\begin{array}{cc} \frac12(1+ \gamma){\cal A}& \frac12(1+ \gamma){\cal B}\\
\frac12(1- \gamma){\cal B}&\ \frac12(1- \gamma){\cal A}\end{array}\right]\equiv 
\left[\begin{array}{cc} A&\ B\\
\widetilde B& \widetilde A\end{array}\right]\label{Xdecomp}
\ ,\ee
\be P=\left[\begin{array}{cc} \frac12(1+ \gamma){\cal V}& \frac12(1+ \gamma){\cal U}\\
\frac12(1- \gamma){\cal U}&\ \frac12(1- \gamma){\cal V}\end{array}\right]\equiv 
\left[\begin{array}{cc} V&\  U \\
\widetilde U& \widetilde V\end{array}\right]\label{Xdecompbis}
\ ,\ee
the action assumes the form
\be\boxed{
\begin{array}{lcl}
S&=&\int_{\cal M} {\rm Tr}_{{{\cal H}}\otimes{\cal F}_0}\Big[ V\star\left( F -  B \star\wt B  
+\tfrac13\, V\star V + U \star\wt U
\right) +
\wt U \star DB  
\\
&&  \ \ \ \ \ \ \ +\wt V\star \left( \wt F - \wt B\star B   
+ \tfrac13\, {\wt V}\star \wt{V}  + \wt U\star U\right) +
U \star \wt D \wt B  \Big]\end{array}}\label{SH1} \ .
\ee
This result for the 3-graded models is to be compared with the action for the $\mathbb Z_2$-graded model 
given in \eq{n3}. The equations of motion at $\partial {\cal M}$, \emph{viz.}
$dX+hXh\star X\approx 0$, resulting from the action above take the form
\be dA+A\star A- B\star \widetilde B\approx0\ ,\qquad 
d \widetilde A+  \widetilde A\star  \widetilde A-  \widetilde B\star  B\approx0\ ,\ee
\be dB+A\star B-B\star \widetilde A\approx0\ ,\qquad d\widetilde B+\widetilde A \star \widetilde B-
\widetilde B \star A\approx 0\ .\ee
In summary, the presence of the extra Klein operator $\gamma$ yields
a refined 3-grading in which ${\cal F}^{(0)}$ is replaced by two blocks, 
namely $\frac12(1\pm \gamma){\cal F}^{(0)}$, where ${\cal F}_0$ is the proper 
Frobenius subalgebra of ${\cal F}$ defined in \eq{calF0}.
The resulting action \eqref{SH1} is of the same form as the original action 
in \eqref{SH} but with more general master fields belonging to $\frac12(1\pm \gamma)$ 
projections of ${{\cal H}}\otimes{\cal F}_0\,$.

\section{Examples}\label{sec:Examples}

In this section we shall provide examples based on unital 
$\mathbb Z_2$-graded Frobenius algebras including 3-graded 
and not 3-graded cases.

\subsection{3-graded matrix algebra}

Unital $\mathbb Z_2$-graded Frobenius algebras with Klein operator $h$
of the form ${\cal F}={\cal F}_0\oplus h{\cal F}_0$ can be obtained by taking
\be {\cal F}_0={\rm mat}_N(\Comp):=\bigoplus_{i,j=1}^N \Comp\otimes m_{i,j}\ ,\qquad
m_{i,j} m_{k,l}:=\delta_{jk} m_{i,l}\ ,\qquad
h \,m_{i,j} \,h := \sigma_i \sigma_j m_{i,j}\ ,\ee
where $\sigma_i\in \{\pm 1\}\,$, and 
\be {\rm Tr}_{\cal F}\, m_{i,j}:=\delta_{i,j}\ ,\qquad {\rm Tr}_{\cal F}\, hm_{i,j}:=0\ .\ee
The decomposition \eqref{calF0} of ${\cal F}_0$ into 
eigenspaces of the adjoint action of $h$ is given by 
\be {\cal F}^+_+=\bigoplus_{i,j=1}^N\Comp \otimes
e_{i,j}\ ,\qquad{\cal F}^-_+=\bigoplus_{i,j=1}^N\Comp \otimes
f_{i,j}\ ,\ee
where     
\be e_{i,j}:=\tfrac12(1+\sigma_i \sigma_j) m_{i,j}\ ,\qquad
f_{i,j}:=\tfrac12(1-\sigma_i \sigma_j) m_{i,j}\ ,\ee
have traces
\be {\rm Tr}_{\cal F} e_{i,j}=\delta_{i,j}\ , \qquad
{\rm Tr}_{\cal F} f_{i,j}=0\ .\ee
The analogous decomposition of $h{\cal F}_0$ leads to the subspaces
\be {\cal F}^+_-=\bigoplus_{i,j=1}^N\Comp\otimes he_{i,j}
\ ,\qquad {\cal F}^-_-=\bigoplus_{i,j=1}^N\Comp\otimes h f_{i,j}\ ,\ee
whose basis elements have traces
\be {\rm Tr}_{\cal F} h\,e_{i,j}=0\ , \qquad{\rm Tr}_{\cal F} h\,f_{i,j}=0\ .\ee
The corresponding master fields 
\be {\cal A}=\sum_{i,j=1}^N  A^{i,j}e_{i,j}\ ,\qquad
{\cal B}=\sum_{i,j=1}^NB^{i,j} f_{i,j}\ ,\ee\be {\cal V}=\sum_{i,j=1}^N V^{i,j} e_{i,j}\ ,\qquad
{\cal U}=\sum_{i,j=1}^N U^{i,j} f_{i,j}\ .\ee
We note that for a given choice of $\sigma_i$ one has
\be N^\sigma= \sum_{i,j=1}^N \tfrac12 (1+\sigma \sigma_i \sigma_j)\ ,\ee
such that if all $\sigma_i$ are equal then the model consists of only
odd forms.

The 3-grading results from the fact that the outer action of 
$h$ on ${\cal F}_0$ is equivalent to the inner adjoint action of 
\be \gamma= \sum_{i=1}^N
\sigma_i m_{i,i}\ ,\label{hgamma}\ee
\emph{viz.} $h m_{i,j} h = \gamma m_{i,j} \gamma$. 
Hence, the above decomposition of ${\cal F}$ can be written as
\be {\cal F}^\pm_+= \tfrac14(1+\gamma) {\cal F}_0(1\pm \gamma)+ 
\tfrac14(1-\gamma) {\cal F}_0(1\mp \gamma) \ .\ee
\be {\cal F}^\pm_-= \tfrac14(1+\gamma) h{\cal F}_0(1\pm \gamma)+ 
\tfrac14(1-\gamma) h{\cal F}_0(1\mp \gamma) \ .\ee
The 3-grading can be used to project the model in order to 
solve the zero-form constraints, as discussed in Section \ref{sec:Three-grading}.
To this end, we permute the basis such that
\be
{\cal F}_0=\left[\begin{array}{cc} {\rm mat}_{N_1}(\Comp) 
& N_1\otimes  N^\ast_2\\ N_2\otimes N_1^\ast&
{\rm mat}_{N_2}(\Comp)\end{array} \right]\cong {\rm mat}_{N}(\Comp)\ ,
\qquad \gamma:=\left[\begin{array}{cc} \mathbb 1_{N_1}&0\\0&-\mathbb 1_{N_2}\end{array} \right]\ ,
\ee
where thus $N=N_1+N_2$, $N^+=(N_1)^2+(N_2)^2$ and $N^-=2N_1 N_2$.
The graded inner product on ${\cal F}$ now reads 
\be (a_0+a'_0h,b_0+b'_0h)_{\cal F}= {\rm Tr}_{{\rm mat}_{N}(\Comp)}  \gamma( a_0 b_0+
a'_0hb'_0h)\ ,\qquad a_0,a'_0,b_0,b'_0\in {\cal F}_0\ ,\ee
where we note that $hb'_0h\in {\cal F}_0$.
The decomposition under the 3-grading now reads 
\be \label{mat1}{\cal F}^{(+0)}= \left[\begin{array}{c|c} {\rm mat}_{N_1}(\Comp) \ 
& \ 0\ \\ \hline 0&0\end{array} \right]\ ,\qquad
{\cal F}^{(+1)}=\left[\begin{array}{c|c} \ 0\ & N_1\otimes  N_2^\ast
                                      \\ \hline 0 &  0\end{array} \right]\ ,
\ee
\be \label{mat2}
{\cal F}^{(-1)}=\left[\begin{array}{c|c}0&\ 0\  \\ \hline N_2\otimes N_1^\ast&0\end{array} \right]\ ,
\qquad {\cal F}^{(-0)}= \left[\begin{array}{c|c} \ 0\ &0\\ \hline 0&
\ {\rm mat}_{N_2}(\Comp)\end{array} \right]\ ,\ee
obeying \eqref{4.28}. 
The resulting model, with action \eqref{SH1} with ${\rm Tr}_{{\cal H}\otimes{\cal F}_0}$
replaced by ${\rm Tr}_{{\cal H}}\,{\rm Tr}_{{\rm mat}_{N}}$,
represents a straightforward extension of the original FCS model 
with $(A,\widetilde A;B,\widetilde B)$ valued in subspaces 
of ${\cal H}\otimes {\rm mat}_{N}(\Comp)$
in accordance with \eqref{mat1} and \eqref{mat2} \emph{idem}
$(V,\widetilde V;U,\widetilde U)$.

\subsection{3-graded Clifford algebra}

For $N=\widetilde N=2^{n-1}$, the 3-graded matrix FCS model introduced 
in the previous section is equivalent to a model with   
\be {\cal F}_0={\cal C}\ell_{2n}\ ,\ee
the Clifford algebra generated by $2n$
elements $\gamma_i$ ($i=1,\dots,2n$) obeying
\be \{\gamma_i,\gamma_j\}=2\delta_{ij}\ .\ee
The trace operation can be defined in the basis consisting
of totally antisymmetric elements 
\be \gamma^{i_1\dots i_p}:= \gamma^{[i_1}\cdots \gamma^{i_p]}\ee
as the projection onto the identity, \emph{i.e.}
\be {\rm Tr}_{{\cal C}\ell_{2n}} \gamma^{i_1\dots i_p}=\delta_{p,0}\ .\ee
The 3-grading is achieved by the inner Klein operator
\be \gamma= i^n \gamma_1\cdots \gamma_{2n}\ .\ee
The resulting model thus consists of odd forms,
not containing top-forms, valued in $\frac14(1+\gamma){\cal C}\ell_{2n}(1+\gamma)$
and $\frac14(1-\gamma){\cal C}\ell_{2n}(1-\gamma)$, both isomorphic to ${\rm mat}_{2^{2n-2}}(\Comp)$,
and even forms, with constrained zero-form and $2n$-form
content, valued in $\frac14(1+\gamma){\cal C}\ell_{2n}(1-\gamma)$
and $\frac14(1-\gamma){\cal C}\ell_{2n}(1+\gamma)$, both
isomorphic to $2^{2n-2}\otimes (2^{2n-2})^\ast$.
In particular, on
\be {\cal M}={\cal X}_5\times {\cal Z}_4\ ,\ee
as in Section~\ref{sec:Section2}, 
it contains a Konstein--Vasiliev phase in which the 
two-form is given by an expectation value 
proportional to the closed and central element
\be J\in \Omega_{[2]}({\cal Z}_4)\otimes {{\cal H}}\ ,\qquad
dJ=0\ ,\qquad J\star f=f\star J\ ,\qquad f\in \Omega({\cal M})\otimes {{\cal H}}\ .
\ee
Fixing gauges for the resulting fluctuations in the forms of
positive degrees is equivalent to performing the consistent 
truncation\footnote{It is important that the truncation
does not affect the zero-form sector.}
\be A=\frac12 (1+\gamma)W\ ,\qquad B=\frac12 (1+\gamma) C \gamma_{2n}\ ,\ee
\be \widetilde B=\frac12(1-\gamma) \gamma_{2n} J\ ,\qquad
 \widetilde A=\frac12 (1-\gamma)\gamma_{2n}W\gamma_{2n}\ ,\ee
where the reduced master fields\footnote{We use a notation in
which ${\rm C}_{{\cal A}}(x)$ denotes the centralizer of an element $x$ 
in an associative algebra ${\cal A}$.} 
\be C,W\in \frac12 (1+\gamma) {\rm C}_{{\cal C}\ell_{2n}}(\gamma)\otimes{{\cal H}}\ ,\qquad
\frac12 (1+\gamma) {\rm C}_{{\cal C}\ell_{2n}}(\gamma)\cong {\rm mat}_{2^{2n-2}}(\Comp)\ , \ee
which yields  
\be dW+W\star W+ C \star J=0\ ,\qquad dC+W\star C-C\star W=0\ .
\ee
Modulo reality and other kinematic conditions\footnote{Whether there 
exist consistent truncations to the Konstein--Vasiliev
models with $husp$ or $ho$ algebras is left for future work.}, 
we identify the above model as an FCS extension of the bosonic Konstein--Vasiliev 
model with gauge algebra $hu(2^{2n-2},0)$ \cite{Konstein:1989ij}.
%

\subsection{Twisted group algebra of $\mathbb{Z}_2\times\mathbb{Z}_{2n}$}

An example of a unital $\mathbb Z_2$-graded Frobenius 
algebra that does not admit any 3-grading is provided by
a twisting of the group algebra\footnote{An outline of twisted group algebras 
is given in Appendix \ref{App:groupalgebras}.} of $\mathbb{Z}_2\times\mathbb{Z}_{2n}$
\emph{viz.}
\be {\cal F}=\Comp[\mathbb{Z}_2\times\mathbb{Z}_{2n},\a]\ ,
\ee
where the group is generated by two elements $a$ and $b$ 
subject to the conditions 
\be
a^{2n} = I = b^2\ ,\qquad  a b = b a\ ,
\label{Z2n}
\ee
and the co-cycle $\alpha$ is chosen such that
\be
e_{a^k}\,e_b = (-1)^{k} e_b \, e_{a^{k}}\ ,\qquad k=0,1,...,2n-1\ .
\ee
As for ${\rm Tr}_{\cal F}$, we take the operation in \eqref{TraceCG}
\footnote{In terms of the basis for $\Comp[\mathbb{Z}_{2n}]$
consisting of the projectors $p_{l}:=\tfrac1{2n} \sum_{k=0}^{2n-1}e^{i\,\tfrac{ kl \pi}{n}} e_{a^k}$, $l=
0,1,\dots,2n-1$, $p_{i} p_{j}=\delta_{i,j}\,p_{i}$, we have 
${\rm Tr}_{\cal F} \,e_i = 1$.
Thus, expanding $x\in \Comp[\mathbb{Z}_2\times\mathbb{Z}_{2n},\a]$ as 
$x= \sum_{l=0}^{2n-1}e_{l} \left(x_{l} + h \, \tilde x_{l}\right)$, $x_{l},\tilde x_{l}\in \Comp$,
its trace ${\rm Tr}_{\cal F}\,x = \sum_{l=0}^{2n-1}x_{l}$.}, and 
the $\mathbb{Z}_2$ grading can be achieved by taking
\be {\cal F}={\cal F}_0\oplus h{\cal F}_0\ ,\qquad  h=e_b\ ,\qquad 
{\cal F}_0=\Comp[\mathbb{Z}_{2n}]\ .\ee
Turning to the master fields, they are given by
\be 
{\cal A}=\sum_{k=1}^n  A^{(2k-2)} e_{a^{2k-2}}\ ,\qquad
{\cal B}=\sum_{k=1}^n B^{(2k-1)} e_{a^{2k-1}}\ ,
\label{exp5}
\ee
\emph{idem} ${\cal V}$ and ${\cal U}\,$.

\paragraph{$\mathbb{Z}_2\times\mathbb{Z}_{4}$ model.} To exhibit 
the structure, let us take $n=2\,$.
The master fields can now be expanded as
\be {\cal A}=\sum_{\sigma=\pm} A_\sigma e_\sigma\ ,\qquad {\cal B}=\sum_{\sigma=\pm} B_\sigma f_\sigma\ ,\ee
\emph{idem} ${\cal U}$ and ${\cal V}$, where the basis elements 
\be e_\sigma = \tfrac12 (e_I +\sigma e_{a^2})\ ,\qquad f_\sigma=e_{a}\,\e_\sigma\ ,\ee
obey
\be e_\sigma e_{\sigma'}=\delta_{\s\s'}e_\sigma\ ,\qquad
e_\sigma f_{\sigma'}=\delta_{\s\s'}f_\sigma
\ ,\qquad
f_\sigma f_{\sigma'}=\sigma \delta_{\s\s'}e_\sigma\ ,\ee
and 
\be {\rm Tr}_{\cal F} e_\sigma=4\ ,\qquad {\rm Tr}_{\cal F} f_\sigma=0\ .\ee
In components, the boundary equations of motion, \emph{viz.} $F_X|_{\partial {\cal M}}= 0$,
with $F_X := dX +hXh \star X = 0$ and $X:={\cal A} + {\cal B}$, read
\begin{eqnarray}
F_\sigma &:=& dA_\sigma+A_\sigma\star A_\sigma \approx \sigma B_\sigma\star B_\sigma\;,\label{new1}\\
D_\sigma  B_\sigma &:=& dB_\sigma +[A_\sigma, B_\sigma]_\star \approx 0\;,\label{new2}
\end{eqnarray}
which is a Cartan integrable system containing the zero-form constraint
\begin{equation}
B_{[0]\sigma}\star B_{[0]\sigma} \approx 0\ .\label{0formc}
\end{equation}
The FCS action \eq{n3} is given by 
\be S=S_+ + S_-\ ,\ee
where
\be S_\sigma=\int_{\cal M}{\rm Tr}_{{\cal H}}\left(
\sigma U_\sigma \star D_\sigma B_\sigma+ V_\sigma \star 
( F_\sigma - \sigma B_\sigma\star B_\sigma + \sigma U_\sigma\star U_\sigma + \tfrac13 V_\sigma\star V_\sigma )\right)\ .\ee
In order for the action to be real and non-degenerate, and the zero-form constraint 
to have a nontrivial solution space, we can impose the reality condition
\be (A_\sigma)^\dagger = - A_{-\sigma} \  ,\qquad    (B_\sigma)^{\dagger} = B_{-\sigma}\ ,\ee
\be (V_\sigma)^\dagger = - V_{-\sigma} \  ,\qquad    (U_\sigma)^\dagger = \sigma U_{-\sigma}\ ,\ee
which implies that $(S_\sigma)^\dagger=S_{-\sigma}$ and that \eqref{new1} and \eqref{new2}
follow from the variational principle.
As for \eqref{0formc}, nontrivial solution spaces arise due to the 
fact that $B_{[0]\sigma}$ is a complex element in ${\cal H}$, 
\emph{e.g.} by using star product realizations of Fock space endomorphisms
\footnote{For a related truncation of three-dimensional fractional spin 
gravity, see Section 4.5 of \cite{Boulanger:2013naa}.}.
\paragraph{Four-dimensional self-dual branch.} Taking 
$\partial{\cal M}_4={\cal X}_4\times {\cal Z}_4$ and
${\cal H}$ to be the four-dimensional bosonic higher 
spin algebra augmented with outer Klein operators $(k,\bar k)$
and using the notation and results of 
\cite{Iazeolla:2007wt,Iazeolla:2011cb,Boulanger:2015kfa},
a branch describing self-dual configurations arises as follows: 
In holomorphic gauge, the constraint \eqref{0formc} is
solved by 
\be B_{[0]+}=\Psi(y,\bar y)\star \kappa_y k\ ,\qquad \Psi\star \Psi=0\ ,\label{psisquared0}\ee
where $\kappa_y=2\pi \delta^2(y)$ is an inner Klein operator and 
$\Psi$ is a nilpotent (complex) Fock space endomorphism 
in the adjoint representation of $hs(4;\Comp)$.
For example, using the star product algebra realization of $|m_+,m_-\rangle\langle n_+,n_-|$
where $m_\pm,n_\pm\in \mathbb Z+\tfrac12$ are eigenvalues of $E\pm J$ (for details, see \cite{Iazeolla:2011cb}), 
one may expand $\Psi$ by taking $m_+=\tfrac12$ mod 4 and $n_+=\tfrac52$ mod 4 (without any need to constrain $m_-,n_-$).
More generally, the solution space of \eqref{psisquared0} decomposes
into $hs(4;\Comp)$ orbits; in this sense, oe may think of $\Psi$
as an higher spin generalization of a pure spinor

The master field equations in positive degrees 
can be solved by setting all forms in degrees greater 
than two to zero, and taking\footnote{Keeping the antiholomorphic component
of the two-form activates the three-form, as $B_{[2]+} \star B_{[2]+}$ is
now proportional to $dz^2 d\bar z^2 \kappa \star \bar\kappa$.}
\be B_{[2]+} = \tfrac{i}8 k \kappa_y\star \kappa_z dz^\alpha dz_\alpha\ ,\qquad A_{[1]+}=dz^\alpha v_{\alpha}(z) \star \Psi
\ee
where the two-form is closed and central and obeys $B_{[2]+}\star B_{[2]+}=0$,
and $v=dz^\alpha v_{\alpha}(z)$ obeys
\be dv=\tfrac{i\pi }2 \delta^2(z) dz^\alpha dz_\alpha\ ,\ee
that can be achieved by taking $v_\alpha$ to have a simple pole at $z^\a=0$.
As shown in \cite{Iazeolla:2011cb}, this singularity can be 
removed by a (unitary) vacuum gauge function\footnote{The 
gauge function is defined on a subset of ${\cal X}_4$.
The sterographic coordinate system $x^\mu\in\Real^{1,3}\setminus \{ x: x^2=1\}$ 
with metric $dx^2/(1-x^2)^2$ covers the coset once.
We take ${\cal X}_4$ to be $\Real^{1,3}$ with points at infinity
such that $\partial{\cal X}_4=0$, and allow the gauge fields 
(but not the curvatures) to blow up on the surface $\{ x: x^2=1\}$.} 
$L:{\cal X}_4\rightarrow SO(2,3)/SO(1,3)$;
the symbol of $L^{-1}\star(A_{[1]+}+d)\star L$ in
Vasiliev's normal order is analytic on ${\cal Z}_4$
(minus the point at infinity)
over a finite region of ${\cal X}_4$.
Thus there exists a field dependent gauge function
that takes the configuration to Vasiliev's gauge,
\emph{i.e.} $z^\alpha A_{\alpha +}=0$ in normal order,
in which vierbein, Lorentz connection, Fronsdal fields 
and Weyl tensors can be defined in a manifestly Lorentz
covariant basis after a field redefinition.

Since the two-form is holomorphic, only the dotted 
Weyl curvatures of the linearized Fronsdal fields 
are sourced by the Weyl zero-form.
Apart from the zero-form constraints and the modified 
reality condition, the mechanism leading to self-dual linearized
curvatures in the current model is the same 
as that spelled out in \cite{Iazeolla:2007wt}\footnote{
In \cite{Iazeolla:2007wt} the zero-form is
unconstrained and proper reality conditions, 
\emph{viz.} $B_{[0]}^\dagger=B_{[0]}$, 
$A_{[1]}^\dagger=-A_{[1]}$ and $\widetilde B_{[2]}^\dagger=-\widetilde B_{[2]}$
are imposed, which requires either (2,2) of (4,0) Lorentz 
signature in order for $\widetilde B_{[2]}$ to be holomorphic.}: 
The calculation of the linearized sources for the Fronsdal 
curvatures follows the same steps as in Vasiliev's original 
work \cite{Vasiliev:1990en}, but since there is no anti-holomorphic 
term in $B_{[2]+}$ the source terms containing 
$\Phi|_{z^\alpha=\bar z^{\ad}=\bar y^{\ad}=0}$ are not present. 
It follows that the linearized curvatures of the Fronsdal fields 
are self-dual, though the zero-forms in $\Phi|_{z^\alpha=\bar z^{\ad}=\bar y^{\ad}=0}$ are 
nonetheless part of the spectrum, playing the role of additional
matter fields.
Thus, the spectrum of dynamical fields on ${\cal X}_4$
consists of a complexified scalar and a tower of self-dual 
complexified gauge fields.
%

\section{Conclusions}\label{sec:Conclusions}

Generalizing the minimal Frobenius--Chern--Simons action of~\cite{Boulanger:2015kfa}, 
we have constructed, under a mild set of assumptions, the most general cubic action 
for a set of even and odd forms on an odd-dimensional manifold
${\cal X}\times {\cal Z}$ where ${\cal X}$ is open and commutative 
and ${\cal Z}$ closed and noncommutative; in the global formulation,
the are on ${\cal X}$ whereas all fields are assumed to be
globally defined on ${\cal Z}$.
The underlying symmetry group is based on the direct product 
${\cal H}\otimes {\cal F}$ of two associative algebras with
non-degenerate invariant inner products.
As for ${\cal H}$, it has been assumed to be unital and trivially graded, which means
that its inner product is a trace operation; in concrete models its role 
is to realize the higher spin algebra and its representations.
The algebra ${\cal F}$, on the other hand, has been assumed to be
finite-dimensional and $\mathbb Z_2$-graded; in the unital case, 
it is thus a $\mathbb Z_2$-graded Frobenius algebra, while we 
refer to it as being quasi-Frobenius in the non-unital case.
In the latter case, the resulting action has the appearance
of a matter-coupled Chern--Simons-like action, as even and odd
forms must be treated on unequal footing.
In the unital case, and under the additional assumption that the $\mathbb Z_2$-grading
can be achieved by an inner Klein operator, the even and odd forms
can be assempled into a single superconnection, resulting in pure
Chern--Simons-like action, that we refer to as a Frobenius--Chern--Simons
action.

Furthermore, we have distinguished between ${\mathbb Z}_2$-graded Frobenius 
algebras and 3-graded versions, and shown that in the latter case constraints 
on zero-form master fields can be avoided. 
In particular, the original model of \cite{Boulanger:2015kfa}, which
admits a perturbative description in terms of real Fronsdal tensors
in $AdS_4$, is based on a 3-graded Frobenius algebra given by a 
twisting of the group algebra based on $(\mathbb Z_2)^3$.
As a simple modification of it, we have shown that a twisting
of the group algebra of $\mathbb Z_2\times \mathbb Z_4$ yields 
a $\mathbb Z_2$-graded Frobenius algebra that leads to a model
with zero-form constraints that admits a perturbative description 
in terms of self-dual complex Fronsdal tensors in $AdS_4$.
Another class of models arise from the 3-graded matrix algebras.
A special case are the Clifford algebras, which lead to an interesting 
off-shell extension of a bosonic subclass of the Konstein-Vasiliev models,
namely those that accommodate the Clifford algebras as an internal symmetry.

In view of the above result and the fact that the four-dimensional higher
spin algebra can be obtained by twisting the algebra of the group 
$SO(2,3)\times {\cal K}$, where ${\cal K}\cong (\mathbb Z_2)^2$,
and factoring out ideals, it would be interesting to undertake 
a more thorough investigation of models based on twisted group 
algebras.
Clearly, many Frobenius algebras may lead to novel equations of 
motion that are not necessarily interpretable as ordinary higher 
spin field equations.
Instead it should be emphasised that the generalized Frobenius--Chern--Simons 
gauge theory presented here may have applications beyond higher spin gravity.

There are several directions for future investigations. 
As already mentioned, it would be interesting to seek new examples 
of Frobenius algebras that lead to novel spectral properties 
and interactions in the context of higher spin gravity, 
compared to the ones known until now \cite{Vasiliev:1992av,Konstein:1989ij,Vasiliev:2003ev,Sezgin:2012ag,Boulanger:2013naa}. 
Of considerable interest are also generalizations of 
Frobenius-Chern-Simons gauge theory that includes 
higher than cubic interactions as well as quadratic
terms. 
Polynomial interactions, including quiver-like 
interactions will be presented elsewhere. 
Finally, it is of great importance to establish a 
connection between such Frobenius-Chern-Simons 
gauge theories and topological open string 
theories \cite{Arias:2015wha,Bonezzi:2015lfa},  
possibly by generalizing the equivalence of 
ordinary Chern-Simons theory
and topological open strings found a long 
time ago by Witten \cite{Witten:1992fb}.
To this end, the addition of quadratic and quartic 
and higher terms to the Hamiltonian
can be shown to lead to extension of 
${\cal F}$ into an internal $A_\infty$ algebra, 
as we hope to report on elsewhere.

Many twisted group algebras are nontrivial viewed as
Hopf algebras, that is, they are co-noncommutative. 
In this respect, it is interesting to note that Hopf 
algebras in the form of quantum groups provide examples 
of differential Poisson manifolds with nontrivial curvatures
that give rise to noncommutative geometries with graded 
non-anticommuting line elements \cite{Beggs:2003ne,McCurdy:2009xz}. 
These types of constructions may give rise to an
even larger landscape of higher spin gravities
provided that one is willing to deform the 
anti-de Sitter symmetry algebra, as makes
sense for example in the application to 
nonrelativitic holographic dualities and in 
particular massive anyons.
More generally, beyond the realm of differential Poisson manifolds reside the homotopy
Poisson manifolds, whose quantization gives rise to a deformation of the external
differential graded algera $\Omega({\cal M})\otimes {\cal H}$ by an external $A_\infty$ algebra.
When combined with the aforementioned internal 
$A_\infty$ algebra, one is thus led to 
a topological version of the category of open 
string field theories proposed by
Gaberdiel and Zwiebach in \cite{Gaberdiel:1997ia}.


\paragraph*{Acknowledgements.}

We thank Pierre Bieliavski and Valentin Ovsienko for 
discussions related to Frobenius algebras.
We are very grateful to Maja Volkov for enlightening discussions 
on extensions of group algebras. 
We also value related and stimulating collaborations 
with Cesar Arias, Carlo Iazeolla and Alexander Torres-Gomez. 
N.B. is thankful to the IH\'ES (Bures-sur-Yvette) and the AEI (Potsdam) 
for hospitality and providing an excellent environment for research.  
The work of E.S. is supported in part by NSF grant PHY-1214344.
and PHY-1521099.
P.S. is grateful to Texas A\&M University and the University of Mons  
for hospitality during various stages of this work.
The work of P.S. is supported by Fondecyt Regular grant N$^{\rm o}$
1140296 and Conicyt grant DPI 20140115 and UNAB grant DI-1382-16/R.

\begin{appendix}

\section{Twisted group algebras}\label{App:groupalgebras}

Given a discrete group $G\,$, the twisted group 
algebra\footnote{Twisted group algebras 
are also known as extensions of standard group algebras 
by an abelian group so that they are group algebras 
of extended groups; for example, p. 31 in \emph{e.g.} 
\cite{serre2005groupes}.} 
\cite{conlon1964twisted}
\be \mathbb C [G,\alpha]=\bigoplus_{g\in G} \Comp 
\otimes e_g\ ,\ee
is the associative algebra with composition rule
\be e_g  e_{g'}=\alpha (g,g') e_{gg'}\ ,\ee
where $\alpha: G\times G\rightarrow \Comp \setminus \{0\}$ is a 
cocycle map.
Associativity implies 
\be \alpha (g,g') \alpha (gg',g'')=\alpha (g,g'g'')\, \alpha (g',g'')\ ,\ee
while the freedom in rescaling $e_g\rightarrow \beta(g) e_g$ 
by nonzero complex numbers $\beta(g)$ implies that the cocycles 
are defined modulo
\be \alpha(g,g')\rightarrow \beta(g) \beta(g') \alpha (g,g') (\beta(gg'))^{-1}\ ,\qquad
\beta:G\rightarrow {\mathbb C}\setminus \{0\}\ .\ee
From $e_g (e_{I} e_{g'})= (e_g e_{I}) e_{g'}$, where $I$ denotes
the identity of $G$, it follows that $\alpha(g,I)=\alpha(I,g')$.
Hence, by making use of the freedom in $\beta(I)$ one can take
\be \alpha(g,I)=\alpha(I,g)=1\ ,\ee
so that $e_I$ becomes the identity in the twisted group algebra,
\emph{i.e.}
\be e_I\, a=a\,e_I= a \quad \mbox{for all $a\in
\mathbb C [G,\alpha]$\ .}\ee
Moreover, from $e_g (e_{g^{-1}} e_{g})= (e_g e_{g^{-1}}) e_{g}$ it follows that
\be \alpha(g,g^{-1})=\alpha(g^{-1},g)\ .\ee
Thus, the twisted group algebra admits the trace operation 
\be {\rm Tr}_{\mathbb C[G,\alpha]} e_g:= \vert G\vert \, \delta_{I,g}\ ,
\quad {\text{where}}\quad \vert G\vert = dim(G)\ .
\label{TraceCG}\ee
Alternatively, the algebra $\mathbb C[G,\alpha]$ can
be thought of as a non-commutative deformation of the 
algebra of $\Comp$-valued functions on $G$, by defining 
a map $V$ that sends $\psi: G\rightarrow \Comp$ to 
\be  V_\psi:=\sum_{g\in G} \psi(g) e_g\ ,\ee
such that 
\be 
V_\psi  V_{\psi'}\equiv  V_{\psi\star \psi'}\ ,
\ee
where the associative star product is given by
\be 
(\psi\star \psi') (g')=\sum_{g\in G} \psi(g) \a(g,g^{-1}g')  \psi(g^{-1}g')  \ .\label{discretestar}
\ee
In this basis, the trace operation is given by 
evaluation at the identity $I\in G$, \emph{viz.}
\be {\rm Tr}_{\mathbb C[G,\alpha]} V_\psi= |G|\psi(I)\ .\ee
If the twisted group algebra is $\mathbb Z_2$-graded by means 
of an \emph{inner} Klein operator $k\,$, then we may either 
take $h=k$ and ${\cal F}=\mathbb C[G,\alpha]$ as in 
Section \ref{sec:UnitalAlgebra}, or $\gamma=k$ and 
${\cal F}_0=\mathbb C[G,\alpha]$ as in Section \ref{sec:Three-grading}.
The discrete groups of order 4 and 8 are
\begin{eqnarray}
G_4 & : & \quad \mathbb{Z}_2 \times \mathbb{Z}_2\ ,\quad  \mathbb{Z}_4
\nonumber \\
G_8 & : & \quad (\mathbb{Z}_2)^3 \ ,\quad \mathbb{Z}_2 \times \mathbb{Z}_4\ ,\quad \mathbb{Z}_8\ ,\quad
\mathbb{D}_4\ ,\quad \mathbb{Q}_8
\end{eqnarray}
The eight-dimensional 3-graded Frobenius algebra introduced in Section 2 is 
isomorphic to $\mathbb C[(\mathbb{Z}_2)^3;\alpha]$ with cocycle factor 
as follows:
Denoting the generators of $(\mathbb{Z}_2)^3$ by $\gamma_i$, $i=1,2,3$,  
obeying the group relations
\be (\gamma_i)^2=I\ ,\qquad \gamma_i\gamma_j=\gamma_j\gamma_i\ ,\ee
where $I$ is the identity element, the subalgebra ${\cal F}_0$ 
can be spanned by
\begin{equation}
e+\tilde e=e_I\;,\quad e-\tilde e=e_{\gamma_1\gamma_2}\;,\quad f+\tilde f=e_{\gamma_1\gamma_3}\;,\quad f-\tilde f=e_{\gamma_2\gamma_3}
\end{equation}
provided $\alpha$ is chosen such that
\be
e_{\gamma_1\gamma_2}\cdot e_{\gamma_1\gamma_2}= e_{\gamma_1\gamma_3}\cdot e_{\gamma_1\gamma_3}=-  
e_{\gamma_2\gamma_3}\cdot e_{\gamma_2\gamma_3}= e_I\ ,\ee
and 
\bea
e_{\gamma_1\gamma_2}\cdot e_{\gamma_1\gamma_3}
&=&e_{\gamma_2\gamma_3}
\;=\; -\, e_{\gamma_1\gamma_3}\cdot e_{\gamma_1\gamma_2} \ , \\[2mm]
\quad e_{\gamma_1\gamma_2}\cdot e_{\gamma_2\gamma_3}
&=& e_{\gamma_1\gamma_3}\;=\;
-\, e_{\gamma_2\gamma_3}\cdot e_{\gamma_1\gamma_2}\ , \\[2mm]
e_{\gamma_2\gamma_3}\cdot e_{\gamma_1\gamma_3}
&\,=\,& e_{\gamma_1\gamma_2} \;=\; 
-\,e_{\gamma_2\gamma_3}\cdot 
e_{\gamma_1\gamma_3} \ .
\end{eqnarray}
The Klein operator
\be h=e_{\gamma_1\gamma_2\gamma_3}\ ,\ee
and we identify
\begin{equation}
h(e-\tilde e)=e_{\gamma_3}\;,
\qquad h(f+\tilde f)=e_{\gamma_2}\;,\qquad h(f-\tilde f)=e_{\gamma_1}\;.
\end{equation}
It follows that the trace operation \eqref{TraceCG} is equivalent to the
trace used in Section 2.
Alternatively, the above algebra can be viewed as the twisted product 
of ${\mathbb Z}_2\times {\mathbb Z}_2$ with another ${\mathbb Z}_2$. 
It would be interesting to determine whether there exist further
twistings leading to nonequivalent FCS models.

The star product \eqref{discretestar} is the discrete counterpart of
the Poincar\'e$\,$--Birkhoff--Witt (PBW) star product on the enveloping
algebra of the Lie algebra $\mathfrak{g}$ of a Lie group $G$:
If $\psi:G\rightarrow \Comp$ then 
$V_\psi=\int_{g\in G} d\mu(g) \psi(g) e_g\in\mathbb C[G] $,
defined using the Haar measure,
can be mapped via $\phi(e_g)=\exp_\star (i 
\phi^\a(g) T_\a)\in {\rm Env}(\mathfrak{g})$,
where $T_\a$ are a set of generators of $\mathfrak{g}$ 
and $\star$ is the PBW product,
to an element $\phi(V_\psi)=\int_{g\in G} d\mu(g) \psi(g) \exp_\star (i 
\phi^\a(g) T_\a)\in {\rm Env}(\mathfrak{g})\,$.
Thus, the basic FCS model in Section \ref{sec:Section2} 
has an internal algebra ${\cal A}={\cal H}\otimes {\cal F}$ 
given by the direct product of two twisted group algebras,
\emph{viz.} the finite-dimensional factor
${\cal F}=\mathbb{C}[(\mathbb{Z}_2)^3;\alpha]$ and 
the infinite-dimensional factor
${\cal H}= {\cal K}\otimes_{\alpha'} {\rm Env}(\mathfrak{so}(3,2))/{\cal I}
\cong\Comp[{\cal K} \times SO(2,3);\alpha']/{\cal I}$, where $\alpha'$ 
encodes the (anti)commutation relations between the outer Klein
operators $k$ and $\bar k$ generating ${\cal K}\cong (\mathbb{Z}_2)^2$
and the generators of $\mathfrak{so}(3,2)$, and $\cal I$ is the singleton 
annihilator.
This suggests that higher spin gravity can be developed further by 
considering internal algebras given by more general infinite-dimensional 
twisted group algebras.

\end{appendix}


\providecommand{\href}[2]{#2}\begingroup\raggedright\endgroup



\end{document}